\documentclass[11pt,epsf,letterpaper]{article}%
\usepackage{color}
\usepackage{amsmath}
\usepackage{amsfonts}
\usepackage{verbatim}
\usepackage{amssymb}
\usepackage{graphicx}
\usepackage{epstopdf}
\usepackage{mathrsfs}
\usepackage{epsfig}
\usepackage{cite}%
\providecommand{\U}[1]{\protect\rule{.1in}{.1in}}
\begin{document}

\title{A Class of Integrable Metrics}
\author{Andr\'{e}s Anabal\'{o}n$^{(1)}$ and Carlos Batista$^{(2)}$\\\textit{$^{(1)}$Departamento de Ciencias, Facultad de Artes Liberales y}\\\textit{Facultad de Ingenier\'{\i}a y Ciencias, Universidad Adolfo
Ib\'{a}\~{n}ez,}\\\textit{Av. Padre Hurtado 750, Vi\~{n}a del Mar, Chile}\\\textit{$^{(2)}$Departamento de F\'\i sica, Universidade Federal de
Pernambuco,} \\\textit{50670-901 Recife-PE, Brazil} \\andres.anabalon@uai.cl, carlosbatistas@df.ufpe.br}
\maketitle

\begin{abstract}
In four dimensions, the most general metric admitting two Killing vectors and
a rank-two Killing tensor can be parameterized by ten arbitrary functions of a
single variable. We show that picking a special vierbien, reducing the system
to eight functions, implies the existence of two geodesic and share-free, null
congruences, generated by two principal null directions of the Weyl tensor.
Thus, if the spacetime is an Einstein manifold, the Goldberg-Sachs theorem
implies it is Petrov type D, and by explicit construction, is in the Carter
class. Hence, our analysis provide an straightforward connection between the
most general integrable structure and the Carter family of spacetimes.

\end{abstract}

\section{Introduction and Discussion}

The Kerr-(A)dS solution was discovered by Carter imposing the Einstein
equations with a cosmological constant on a family of metrics identified by
the requirement of the separability of the Schr\"{o}dinger and Hamilton-Jacobi
equations \cite{Carter-KleinG}. The mathematical structure behind the
Hamilton-Jacobi separability on a spacetime with two Killing vectors is the
existence of a Killing tensor (a modern review on the subject can be found in
\cite{Kubiznak:2008qp}). The most general $D$-dimensional metric with a
rank-two Killing tensor and $D-2$ Killing vectors was found by Benenti and
Francaviglia \cite{BenentiFrancaviglia}. In the same paper, it is pointed out
that the requirement of Schr\"{o}dinger separability is redundant and that the
new metrics contain the Carter metric as a special subcase.

The construction of the Carter form of the metric is heuristically explained
in the lectures given in the \textquotedblleft Les Houches Ecole d'Et\'{e} de
Physique Th\'{e}orique\textquotedblright\ \cite{DeWitt:1973uma}. Requiring the
separability of the Klein-Gordon equation with a mass term, Carter ends up
with an inverse metric of the form\footnote{see eqs. (5.10)-(5.18) in
\cite{DeWitt:1973uma}}:%

\begin{equation}
\left(  \frac{d}{ds}\right)  ^{2}=\frac{1}{Z}\left\{  \Delta_{x}\left(
\partial_{x}\right)  ^{2}+\Delta_{y}\left(  \partial_{y}\right)  ^{2}+\frac
{1}{\Delta_{x}}\left[  Z_{x}\left(  \partial_{t}\right)  +C_{x}\left(
\partial_{\varphi}\right)  \right]  ^{2}-\frac{1}{\Delta_{y}}\left[
Z_{y}\left(  \partial_{t}\right)  +C_{y}\left(  \partial_{\varphi}\right)
\right]  ^{2}\right\}  \text{ ,}\label{CA}%
\end{equation}
with $Z=C_{y}Z_{x}-C_{x}Z_{y}$. The metric depends on four arbitrary functions
of the coordinates $(x,y)$, namely $\left\{  Z_{y},Z_{x},\Delta_{y},\Delta
_{x}\right\}  $. $C_{x}$ and $C_{y}$ are constants. Indeed, the Carter ansatz
(\ref{CA}) is a special case of the Benenti-Francaviglia metric:
\begin{equation}
g^{ab}\partial_{a}\partial_{b}=\frac{1}{S_{1}(x)+S_{2}(y)}\,\left[  \left(
\,F_{1}^{ij}(x)\,-\,F_{2}^{ij}(y)\,\right)  \partial_{i}\partial_{j}%
+\Delta_{1}(x)\,\left(  \partial_{x}\right)  ^{2}+\Delta_{2}(y)\,\left(
\partial_{y}\right)  ^{2}\right]  \,\text{\ ,}\label{metric1}%
\end{equation}
where the indices $a,b$ range over the coordinates $\{\tau,\sigma,x,y\}$, and
the indices $i,j$ run over $\{\tau,\sigma\}$. $\left\{  F_{1}^{ij}=F_{1}%
^{ji},S_{1},\Delta_{i}\right\}  $ are five arbitrary functions depending on
$x$ and $\left\{  F_{1}^{ij}=F_{1}^{ji},S_{1},\Delta_{i}\right\}  $ are five
arbitrary functions depending on $y$. In this paper we show, by explicit
calculation, that the system of equations of eight arbitrary functions,
generated by replacing (\ref{metric1}) in the Einstein equations with a
cosmological constant can be fully integrated requiring only that:
\begin{equation}
F_{1}^{\tau\sigma}\,=\,\sqrt{F_{1}^{\tau\tau}F_{1}^{\sigma\sigma}}\quad\text{
and }\quad F_{2}^{\tau\sigma}=\sqrt{F_{2}^{\tau\tau}F_{2}^{\sigma\sigma}%
}\,.\label{2}%
\end{equation}
Furthermore, special attention is given to the existence of Killing-Yano
(\textbf{KY}) tensors. In particular, we explicitly show how these tensors can
be quite helpful in the integration of Einstein's equation. Moreover, KY
tensors have proved to be a valuable tool in the study of black holes. Indeed,
the analytic integration of the geodesic equation \cite{Carter-KleinG} as well
as the Klein-Gordon and Dirac equations \cite{Chandra-Dirac} in 4-dimensional
Kerr spacetime is possible due to the existence of an integration constant
constructed using the non-trivial Killing tensor of order two
\cite{Carter-constant,Walk-Pen}. Since this Killing tensor is the square of a
KY tensor of order two \cite{Collinson-Stephani}, the integrability can be
traced to the existence of a KY tensor. Likewise, KY tensors have proved to
play a central role on the integrability of higher-dimensional black holes
\cite{Frolov_KY,Yasui}. Indeed, the class of Kerr-NUT-(A)dS spacetimes in
arbitrary dimension admits a tower of KY tensors that enables the analytical
integration of the geodesic equation \cite{Kubiz,Krtous} along with the
Klein-Gordon \cite{Frol-KG}, and Dirac equations \cite{Oota} in this
background. KY tensors are also related to the separability of gravitational
perturbations in these black holes \cite{Teukolsky,OotaPerturb}.

The interplay between supersymmetry and KY tensors have been discussed in the
literature, see for instance \cite{KY-SUSY}. Moreover, the Carter form of the
metric has been used as the seed to find spinning solutions in gauged and
ungauged supergravity \cite{Chong:2004na}. The same form of the metric has
been used to study the existence of supersymmetric solutions
\cite{AlonsoAlberca:2000cs, Klemm:2013eca}. The more general class
(\ref{metric1}), fits, in the string frame, the large family of rotating black
holes which were recently found in the $U(1)^{4}$ invariant sector of gauged
$\mathcal{N}=8$ supergravity in four dimensions \cite{Chow:2010sf,
Chow:2010fw, Chow:2013tia, Chow:2013gba, Chow:2014cca}. When multiplied by an
arbitrary conformal factor, the metric (\ref{CA}) has been shown to be
integrable in the presence of a real scalar field with an arbitrary scalar
field self-interaction; the scalar field potential being integrated a
posteriori and singled out by the form of the metric \cite{Anabalon:2012ta}.

Therefore, it is worth to have at hand a systematic analysis of these
ans\"{a}tze in general cases. The results of section two, thus provide the
conformal properties of the metric (\ref{metric1}) under the condition
(\ref{2}). Namely, without imposing a field equation on the metric, the
existence of geodesic and share-free null-congruences is established by
explicit construction. Later, using these conformal properties and the
Goldberg-Sach theorem we impose the Petrov type D condition on the metric. The
remaining of the paper is dedicated to the integration of the Einstein
equations with a cosmological constant, trying to be exhaustive in the
analysis of subcases and existence of peculiar geometrical structures in every case.

The whole process can be done in the presence of a real scalar field with an
arbitrary self-interaction along the lines of \cite{Anabalon:2012ta}. In this
case, we found that the metric has to be conformally flat and that the scalar
field and the spacetime are singular, we do not give the details of this
result here. The paper leaves the door open to follow the study of the metric
(\ref{metric1}) without the condition (\ref{2}). This is particularly
interesting in the case when the cosmological constant is non-zero. No
uniqueness theorem for the rotating black holes exists for asymptotically
(anti)-de Sitter spacetimes \cite{Robinson:2004zz}.

\section{The Conformal Properties of the metric}

\label{Sec.General}

If a 4-dimensional spacetime posses just two independent Killing vector fields
then one can build three first integral, two from the explicit symmetries and
one by the metric, which is a Killing tensor. Generally, these three constants
of motion are not enough for an analytical integration of the geodesic
equation. Nevertheless, if, besides the two Killing vectors and the metric,
the spacetime has a non-trivial Killing tensor then one can build one extra
first integral and the integration by quadratures of the geodesics is indeed
possible. Moreover, this extra symmetry can also lead to the integrability of
the Klein-Gordon and Dirac field equations in these backgrounds, as it happens
to be the case with Kerr metric \cite{Carter-KleinG,Chandra-Dirac}. Hence, we
shall study the class of metrics with two commuting Killing vectors and one
non-trivial Killing tensor of order two (\ref{metric1}):
\begin{equation}
\boldsymbol{K}\,=\,\frac{-1}{S_{1}+S_{2}}\,\left[  \left(  \,F_{1}^{ij}%
\,S_{2}+S_{1}\,F_{2}^{ij}\,\right)  \partial_{i}\partial_{j}+\Delta_{1}%
\,S_{2}\left(  \partial_{x}\right)  ^{2}-S_{1}\,\Delta_{2}\left(  \partial
_{y}\right)  ^{2}\right]  \,. \label{KillingT1}%
\end{equation}
As we mentioned in the introduction, we will focus in the particular case
where the following, degenerated vierbein exists:
\begin{equation}
F_{1}^{ij}\,\partial_{i}\partial_{j}\,=\,\left[  \,f_{1}(x)\,\partial_{\tau
}\,+\,h_{1}(x)\,\partial_{\sigma}\,\right]  ^{2}\quad,\quad F_{2}%
^{ij}\,\partial_{i}\partial_{j}\,=\,\left[  \,f_{2}(y)\,\partial_{\tau
}\,+\,h_{2}(y)\,\partial_{\sigma}\,\right]  ^{2}\,.
\end{equation}
Then, defining $S(x,y)=S_{1}(x)+S_{2}(y)$ along with the vector fields
\begin{align}
\boldsymbol{l}  &  =\frac{1}{\sqrt{2\,S}}\left[  \,f_{2}\,\partial_{\tau
}+\,h_{2}\,\partial_{\sigma}+\,\sqrt{\Delta_{2}}\,\partial_{y}\right]  \,,\\
\boldsymbol{n}  &  =\frac{1}{\sqrt{2\,S}}\left[  \,f_{2}\,\partial_{\tau
}+\,h_{2}\,\partial_{\sigma}-\,\sqrt{\Delta_{2}}\,\partial_{y}\right]  \,,\\
\boldsymbol{m}_{1}  &  =\frac{1}{\sqrt{2\,S}}\left[  \,f_{1}\,\partial_{\tau
}+\,h_{1}\,\partial_{\sigma}+i\,\,\sqrt{\Delta_{1}}\,\partial_{x}\right]
\,,\label{NullTetrad1}\\
\boldsymbol{m}_{2}  &  =\frac{1}{\sqrt{2\,S}}\left[  \,f_{1}\,\partial_{\tau
}+\,h_{1}\,\partial_{\sigma}-i\,\,\sqrt{\Delta_{1}}\,\partial_{x}\right]
\end{align}
we have that the metric can be written as\footnote{In what follows, the symbol
$\odot$ stands for the symmetrized tensorial product of two vector fields. For
instance, $\boldsymbol{l}\odot\boldsymbol{n}\,=\,\boldsymbol{l}\otimes
\boldsymbol{n}+\boldsymbol{l}\otimes\boldsymbol{n}$.}:
\begin{equation}
\boldsymbol{g}\,=\,-\,\boldsymbol{l}\odot\boldsymbol{n}\,+\,\boldsymbol{m}%
_{1}\odot\boldsymbol{m}_{2}\,. \label{g-nullTetrad}%
\end{equation}
So, we have that $\{\boldsymbol{l},\boldsymbol{n},\boldsymbol{m}%
_{1},\boldsymbol{m}_{2}\}$ is a null tetrad, namely the only non-vanishing
inner products between the vectors of this basis are:
\[
l^{a}\,n_{a}\,=\,-1\quad\text{ and }\quad m_{1}^{\,a}\,m_{2\,a}\,=\,1\,.
\]
Using this frame, the Killing tensor (\ref{KillingT1}) can be conveniently
written as:
\begin{equation}
\boldsymbol{K}\,=\,-\,S_{1}(x)\,\boldsymbol{l}\odot\boldsymbol{n}%
\,-\,S_{2}(y)\,\boldsymbol{m}_{1}\odot\boldsymbol{m}_{2}\,. \label{KillingT2}%
\end{equation}
The nice thing about this null tetrad is that, if we use the metric to
transform the vector fields $\{\boldsymbol{l},\boldsymbol{n},\boldsymbol{m}%
_{1},\boldsymbol{m}_{2}\}$ into 1-forms, one can check that the following
relations hold:
\begin{align}
d\boldsymbol{l}\wedge\boldsymbol{l}\wedge\boldsymbol{m}_{1}  &  =0\quad\text{
and }\quad d\boldsymbol{m}_{1}\wedge\boldsymbol{l}\wedge\boldsymbol{m}%
_{1}=0\,,\\
d\boldsymbol{l}\wedge\boldsymbol{l}\wedge\boldsymbol{m}_{2}  &  =0\quad\text{
and }\quad d\boldsymbol{m}_{2}\wedge\boldsymbol{l}\wedge\boldsymbol{m}%
_{2}=0\,,\\
d\boldsymbol{n}\wedge\boldsymbol{n}\wedge\boldsymbol{m}_{1}  &  =0\quad\text{
and }\quad d\boldsymbol{m}_{1}\wedge\boldsymbol{n}\wedge\boldsymbol{m}%
_{1}=0\,,\label{IntegrableDist}\\
d\boldsymbol{n}\wedge\boldsymbol{n}\wedge\boldsymbol{m}_{2}  &  =0\quad\text{
and }\quad d\boldsymbol{m}_{2}\wedge\boldsymbol{n}\wedge\boldsymbol{m}%
_{2}=0\,.
\end{align}
According to the Frobenius theorem, the first of these four relations
guarantees that the surfaces orthogonal to $Span\{\boldsymbol{l}%
,\boldsymbol{m}_{1}\}$ form a locally integrable foliation of the manifold.
However, since the vectors $\{\boldsymbol{l},\boldsymbol{m}_{1}\}$ are both
null and orthogonal to each other, it follows that these orthogonal surfaces
are tangent to $Span\{\boldsymbol{l},\boldsymbol{m}_{1}\}$ itself. Since the
tangent vectors to these surfaces are all null, and the maximum dimension of a
null subspace in four dimensions is two, we say that $Span\{\boldsymbol{l}%
,\boldsymbol{m}_{1}\}$ is a maximally isotropic integrable distribution.
Analogously, the three remaining relations in (\ref{IntegrableDist}) imply
that $Span\{\boldsymbol{l},\boldsymbol{m}_{2}\}$, $Span\{\boldsymbol{n}%
,\boldsymbol{m}_{1}\}$ and $Span\{\boldsymbol{n},\boldsymbol{m}_{2}\}$ are
also maximally isotropic integrable distributions. In the Lorentzian case this
is tantamount to saying that the vector fields $\boldsymbol{l}$ and
$\boldsymbol{n}$ are both geodesic and shear-free \cite{RobinsonManifolds}.

Now, without loss of generality, one can write the functions $f_{1}%
,h_{1},f_{2}$ and $h_{2}$ as follows:
\begin{equation}
f_{1}(x)\,=\,\frac{-\,P_{1}(x)}{\sqrt{A_{1}(x)\,\Delta_{1}(x)}}\,\quad,\quad
h_{1}(x)\,=\,\frac{1}{\sqrt{A_{1}(x)\,\Delta_{1}(x)}}\,,
\end{equation}%
\begin{equation}
f_{2}(y)\,=\,\frac{P_{2}(y)}{\sqrt{A_{2}(y)\,\Delta_{2}(y)}}\,\quad,\quad
h_{2}(y)\,=\,\frac{1}{\sqrt{A_{2}(y)\,\Delta_{2}(y)}}\,.
\end{equation}
With these definitions the metric (\ref{metric1}) has the following line
element:
\begin{equation}
ds^{2}=S\,\left[  \frac{-\,A_{2}\,\Delta_{2}}{(P_{1}+P_{2})^{2}}\,\left(
\,d\tau+P_{1}\,d\sigma\,\right)  ^{2}\,+\,\frac{A_{1}\,\Delta_{1}}%
{(P_{1}+P_{2})^{2}}\,\left(  \,d\tau-P_{2}\,d\sigma\,\right)  ^{2}%
\,+\,\frac{dx^{2}}{\Delta_{1}}\,+\,\frac{dy^{2}}{\Delta_{2}}\right]  .
\label{LineEl.P}%
\end{equation}
In order to integrate Einstein's equation for the metric (\ref{LineEl.P}), it
is useful to take advantage of the Goldberg-Sachs theorem. In its original
version \cite{Goldberg-Sachs}, such theorem states that a Ricci-flat
4-dimensional spacetime admits a geodesic and shear-free null congruence if,
and only if, the Weyl tensor is algebraically special according to the Petrov
classification with the repeated principal null direction being tangent to the
shear-free congruence. Since the Weyl tensor, as well as the geodesic and
shear-free property of a null congruence, are invariant under conformal
transformations, it was soon realized that the Ricci-flat hypothesis could be
weakened and replaced by a conformally invariant condition \cite{GS-Conformal}%
. In particular, it was proved that the Goldberg-Sachs theorem also holds in
the presence of a cosmological constant. Later, a version of the
Goldberg-Sachs theorem valid in 4-dimensional manifolds of arbitrary signature
was also proved \cite{Plebanski2}. Particularly, in non-Lorentzian signature
the concept of geodesic and shear-free null congruence might be replaced by an
integrable distribution of isotropic planes. Regarding the metric investigated
here, (\ref{LineEl.P}), we have seen that the null vector fields
$\boldsymbol{l}$ and $\boldsymbol{n}$ are geodesic and shear-free. Therefore,
the Goldberg-Sachs theorem guarantees that whenever $R_{ab}=\Lambda\,g_{ab}$
holds, with $R_{ab}$ denoting the Ricci tensor, $\boldsymbol{l}$ and
$\boldsymbol{n}$ will be repeated principal null directions of the Weyl
tensor. In particular, this means that the Petrov type of the Weyl tensor is
$D$. So, imposing the Weyl tensor of the metric (\ref{LineEl.P}) to be type
$D$ represents no constraint if Einstein's vacuum equation with a cosmological
constant is assumed. Thus, our next step is to impose the type $D$ condition
to the metric (\ref{LineEl.P}).

Denoting the Weyl tensor by $C_{abcd}$, in the Lorentzian signature the
components of the Weyl tensor can be assembled in the following five complex
scalars \cite{Bat1}:
\begin{gather}
\Psi_{0}^{+}\equiv C_{abcd}\,l^{a}m_{1}^{\,b}l^{c}m_{1}^{\,d}\;;\;\Psi_{1}%
^{+}\equiv C_{abcd}\,l^{a}n^{b}l^{c}m_{1}^{\,d}\;;\;\Psi_{2}^{+}\equiv
C_{abcd}\,l^{a}m_{1}^{\,b}m_{2}^{\,c}n^{d}\\
\Psi_{3}^{+}\equiv C_{abcd}\,l^{a}n^{b}m_{2}^{\,c}n^{d}\;;\;\Psi_{4}^{+}\equiv
C_{abcd}\,n^{a}m_{2}^{\,b}n^{c}m_{2}^{\,d}\,. \label{Weylscalars}%
\end{gather}
Computing these scalars for the metric (\ref{LineEl.P}) we find that $\Psi
_{0}=0=\Psi_{4}$, which means that $\boldsymbol{l}$ and $\boldsymbol{n}$ are
principal null directions of the Weyl tensor. In this case, the type $D$
constraint amounts to imposing both $\Psi_{1}$ and $\Psi_{3}$ to vanish.
However, one can check that for this line element the relation $\Psi_{1}%
=\Psi_{3}$ holds, so that we just need to impose $\Psi_{1}$ to vanish. Solving
this constraint, we find that the Petrov type of metric (\ref{LineEl.P}) is
$D$ if, and only if, $A_{1}$ and $A_{2}$ can be put in the following form:
\begin{equation}
A_{1}(x)\,=\,\frac{\left(  \,P_{1}^{\prime}\,\right)  ^{2}}{4\,\left(
\,b_{1}\,P_{1}+\eta_{1}\,\right)  \left(  \,b_{2}\,P_{1}+\eta_{2}\,\right)
}\quad,\quad A_{2}(y)\,=\,\frac{-\,(\,P_{2}^{\prime}\,)^{2}}{4\,\left(
\,b_{1}\,P_{2}-\eta_{1}\,\right)  \left(  \,b_{2}\,P_{2}-\eta_{2}\,\right)
}\,. \label{A1A2-general}%
\end{equation}
Where $P_{1}^{\prime}$ and $P_{2}^{\prime}$ stand for the first derivatives of
$P_{1}(x)$ and $P_{2}(y)$ respectively, whereas $b_{1}$, $b_{2}$, $\eta_{1}$
and $\eta_{2}$ are arbitrary constants. Note that if $b_{2}\neq0$ then one can
always absorb a multiplicative factor in the other constants and make
$b_{2}=1$. In spite of such freedom, for reasons of aesthetic symmetry, we
shall not take advantage of this possibility. It is worth stressing that the
above expressions are not valid if either $P_{1}$ or $P_{2}$ are constant
functions, since in this case $A_{1}$ or $A_{2}$ would vanish according to
(\ref{A1A2-general}), which would imply the determinant of the metric to
vanish. Indeed, one can check that if $P_{1}^{\prime}=0$ and $P_{2}^{\prime
}\neq0$ then, in order for the metric to be type $D$, the function $A_{1}(x)$
can be arbitrary while $A_{2}(y)$ might be given by
\begin{equation}
A_{2}(y)\,=\,\frac{c\,\left(  P_{2}^{\prime}\right)  ^{2}}{\left(
\,P_{1}+P_{2}\,\right)  ^{2}}\quad\;\;(P_{1}\,=\,\text{constant})\,,
\label{A2-P1CteD}%
\end{equation}
with $c$ being a non-zero constant. Analogously, if $P_{2}^{\prime}=0$ and
$P_{1}^{\prime}\neq0$ then, in order for the Petrov classification to be type
$D$, the function $A_{2}(y)$ can be arbitrary while $A_{1}(x)$ might be given
by
\begin{equation}
A_{1}(x)\,=\,\frac{c\,\,\left(  P_{1}^{\prime}\right)  ^{2}}{\left(
\,P_{1}+P_{2}\,\right)  ^{2}}\quad\;\;(P_{2}\,=\,\text{constant})\,.
\label{A1-P2CteD}%
\end{equation}
Finally, if $P_{1}$ and $P_{2}$ are both constant then the metric
(\ref{A1A2-general}) is automatically type $D$. In forthcoming sections
Einstein's equation for the metric (\ref{A1A2-general}) will be fully
integrated and the type $D$ condition will be helpful for the achievement of
this goal. We shall separate our analysis in three cases depending on whether
the functions $P_{1}(x)$ and $P_{2}(y)$ are constant or not.

As an aside, it is worth noting that along these calculations to impose the
type $D$ condition it was implicitly assumed that the signature is Lorentzian.
In the non-Lorentzian case the self-dual and the anti-self-dual parts of the
Weyl tensor are not related to each other by complex conjugation, so that
besides the five Weyl scalars defined in (\ref{Weylscalars}) one must also
consider following other five scalars \cite{Bat1}:
\begin{gather}
\Psi_{0}^{-}\equiv C_{abcd}\,l^{a}m_{2}^{\,b}l^{c}m_{2}^{\,d}\;;\;\Psi_{1}%
^{-}\equiv C_{abcd}\,l^{a}n^{b}l^{c}m_{2}^{\,d}\;;\;\Psi_{2}^{-}\equiv
C_{abcd}\,l^{a}m_{2}^{\,b}m_{1}^{\,c}n^{d}\label{Weylscalars2}\\
\Psi_{3}^{-}\equiv C_{abcd}\,l^{a}n^{b}m_{1}^{\,c}n^{d}\;;\;\Psi_{4}^{-}\equiv
C_{abcd}\,n^{a}m_{1}^{\,b}n^{c}m_{1}^{\,d}\,.
\end{gather}
In spite of this further complication in the non-Lorentzian case, one can
check that the above restrictions for the functions $A_{1}(x)$ and $A_{2}(y)$
also imply that the anti-self-dual part of the Weyl tensor is type $D$, namely
the Weyl scalars $\Psi_{0}^{-}$, $\Psi_{1}^{-}$, $\Psi_{3}^{-}$ and $\Psi
_{4}^{-}$ vanish simultaneously. Thus, for an arbitrary signature, the
conditions (\ref{A1A2-general}), (\ref{A2-P1CteD}) and (\ref{A1-P2CteD}) imply
that the algebraic type of the Weyl tensor is $(D,D)$ according to the
generalized Petrov classification \cite{Bat1}.


\section{Integrating Einstein's Equation for the General Case}

\label{Sec.Einstein-General}

In the present section let us deal with the general case in which $P_{1}(x)$
and $P_{2}(y)$ are both non-constant functions. In this case, one can define
new coordinates $\hat{x}=\sqrt{P_{1}(x)}$ and $\hat{y}=\sqrt{P_{2}(y)}$ and
then judiciously redefine $A_{1},\Delta_{1},A_{2}$ and $\Delta_{2}$ in such a
way that, omitting the hats, the line element (\ref{LineEl.P}) becomes:
\begin{equation}
ds^{2}=S\,\left[  \frac{-\,A_{2}\,\Delta_{2}}{(x^{2}+y^{2})^{2}}\,\left(
\,d\tau+x^{2}\,d\sigma\,\right)  ^{2}\,+\,\frac{A_{1}\,\Delta_{1}}%
{(x^{2}+y^{2})^{2}}\,\left(  \,d\tau-y^{2}\,d\sigma\,\right)  ^{2}%
\,+\,\frac{dx^{2}}{\Delta_{1}}\,+\,\frac{dy^{2}}{\Delta_{2}}\right]  \,.
\label{LineEl.Canonical}%
\end{equation}
The goal of this section is to solve Einstein's vacuum equation for the metric
(\ref{LineEl.Canonical}), namely we shall integrate the equation
\begin{equation}
R_{ab}\,=\,\Lambda\,g_{ab}\,. \label{EinsteinEq}%
\end{equation}
As explained in the preceding section, if (\ref{EinsteinEq}) holds then the
algebraic type of the Weyl tensor for the metric considered here is $D$, so
that the functions $A_{1}$ and $A_{2}$ are given by (\ref{A1A2-general}). In
particular, since we have chosen a gauge in which $P_{1}(x)$ and $P_{2}(y)$
are $x^{2}$ and $y^{2}$ respectively, it follows that
\begin{equation}
A_{1}(x)\,=\,\frac{x^{2}}{(b_{1}\,x^{2}+\eta_{1})(b_{2}\,x^{2}+\eta_{2})}%
\quad\text{and}\quad A_{2}(y)\,=\,\frac{-\,y^{2}}{(b_{1}\,y^{2}-\eta
_{1})(b_{2}\,y^{2}-\eta_{2})}\,. \label{A1A2}%
\end{equation}
It is worth noting that if $S=x^{2}+y^{2}$ and $A_{1}=1=A_{2}$ ($b_{1}%
=0=\eta_{2}$ and $\eta_{1}=1=b_{2}$), the above line element reduces to the
canonical form of Carter's metric \cite{Carter-KleinG}. Particularly, assuming
$S=x^{2}+y^{2}$ along with $A_{1}=1=A_{2}$ and then solving Einstein's vacuum
equation with a cosmological constant we are lead to Kerr-NUT-(A)dS metric
\cite{FrolovBook}. In the present article, we shall go one step further and
integrate Einstein's vacuum equation with a cosmological constant for the full
metric (\ref{LineEl.Canonical}), with $S=S_{1}(x)+S_{2}(y)$, $A_{1}(x)$,
$A_{2}(y)$, $\Delta_{1}(x)$ and $\Delta_{2}(y)$ being, in principle, arbitrary functions.

Einstein's vacuum equation, $R_{\phantom{a}b}^{a}=\Lambda\,\delta
_{\phantom{a}b}^{a}$, implies that $R_{\phantom{x}y}^{x}=0$ which, in turn, is
equivalent to the following differential equation:
\begin{equation}
4\,x\,y\left(  \,S_{1}+S_{2}\,\right)  ^{2}\,=\,\left(  \,x^{2}+y^{2}%
\,\right)  ^{2}\,\frac{dS_{1}}{dx}\,\frac{dS_{2}}{dy}\,. \label{DifEqS}%
\end{equation}
Working out the general solution of (\ref{DifEqS}) yields
\begin{equation}
S_{1}(x)\,=\,\frac{b_{3}\,x^{2}+\eta_{3}}{b_{4}\,x^{2}+\eta_{4}}\quad,\quad
S_{2}(y)\,=\,-\,\frac{b_{3}\,y^{2}-\eta_{3}}{b_{4}\,y^{2}-\eta_{4}}\,.
\label{S1S2-1}%
\end{equation}
Where $b_{3}$, $b_{4}$, $\eta_{3}$ and $\eta_{4}$ are arbitrary constants.
Now, inserting (\ref{A1A2}) and (\ref{S1S2-1}) into the equation $R_{ab}%
l^{a}l^{b}=0$ one can see that one of the following relations must hold:
\begin{equation}
b_{4}\,\eta_{1}\,-\,b_{1}\,\eta_{4}\,=\,0\quad\text{ or }\quad\,b_{4}%
\,\eta_{2}\,-\,b_{2}\,\eta_{4}\,=\,0\,.
\end{equation}
Assuming that $b_{4}\neq0$, we can set $b_{4}=b_{1}$ in (\ref{S1S2-1}) by
redefinition of the other integration constants. Thus, it follows that, up to
a permutation of the integration constants that $b_{4}\neq0$ and $R_{ab}%
l^{a}l^{b}=0$ implies $\eta_{4}=\eta_{1}$. Therefore, the conformal factor
$S_{1}(x)+S_{2}(y)$ is given in terms of the functions
\begin{equation}
S_{1}(x)\,=\,\frac{b_{3}\,x^{2}+\eta_{3}}{b_{1}\,x^{2}+\eta_{1}}\quad,\quad
S_{2}(y)\,=\,-\,\frac{b_{3}\,y^{2}-\eta_{3}}{b_{1}\,y^{2}-\eta_{1}}\,.
\label{S1S2-2}%
\end{equation}
Assuming (\ref{A1A2}) and (\ref{S1S2-2}) to hold we have that the following
eight components of Einstein's vacuum equation are immediately satisfied:
\begin{align}
R_{ab}\,l^{a}l^{b}\,=\,R_{ab}\,n^{a}n^{b}\,=\,R_{ab}\,m_{1}^{\,a}m_{1}%
^{\,b}\,=\,R_{ab}\,m_{2}^{\,a}m_{2}^{\,b}  &  \,=\,0\,,\\
R_{ab}\,l^{a}m_{1}^{\,b}\,=\,R_{ab}\,l^{a}m_{2}^{\,b}\,=\,R_{ab}\,n^{a}%
m_{1}^{\,b}\,=\,R_{ab}\,n^{a}m_{2}^{\,b}  &  \,=\,0\,. \label{Einstein8}%
\end{align}
Hence, it just remains to integrate the equations $R_{ab}\,m_{1}^{\,a}%
m_{2}^{\,b}=\Lambda$ and $R_{ab}\,l^{a}n^{b}=-\Lambda$, which yield a coupled
system of linear differential equations for $\Delta_{1}(x)$ and $\Delta
_{2}(y)$ whose general solution is
\begin{align}
\Delta_{1}(x)\,  &  =\,\frac{I_{1}\,J_{1}}{x^{2}}\,\left[  d_{1}\,I_{1}%
^{3/2}\,J_{1}^{1/2}\,+\,d_{2}\,I_{1}^{2}\,+\,d_{3}\,I_{1}\,J_{1}%
\,+\,\frac{\Lambda}{3b_{1}^{2}}\frac{b_{1}\eta_{3}-b_{3}\eta_{1}}{b_{2}%
\eta_{1}-b_{1}\eta_{2}}\right]  \,,\nonumber\\
&  \quad\label{Delta1Delta2}\\
\Delta_{2}(y)\,  &  =\,\frac{I_{2}\,J_{2}}{y^{2}}\,\left[  d_{4}\,I_{2}%
^{3/2}\,J_{2}^{1/2}\,-\,d_{2}\,I_{2}^{2}\,-\,d_{3}\,I_{2}\,J_{2}%
\,-\,\frac{\Lambda}{3b_{1}^{2}}\frac{b_{1}\eta_{3}-b_{3}\eta_{1}}{b_{2}%
\eta_{1}-b_{1}\eta_{2}}\right]  \,.\nonumber
\end{align}
Where the $d$'s are arbitrary constants and $I_{1}$, $I_{2}$, $J_{1}$ and
$J_{2}$ represent the following functions:
\begin{align}
I_{1}(x)=b_{1}\,x^{2}\,+\,\eta_{1}\quad &  ,\quad J_{1}(x)=b_{2}%
\,x^{2}\,+\,\eta_{2}\,,\\
I_{2}(y)=b_{1}\,y^{2}\,-\,\eta_{1}\quad &  ,\quad J_{2}(y)=b_{2}%
\,y^{2}\,-\,\eta_{2}\,.
\end{align}
Thus, we have completely integrated Einstein's vacuum equations with a
cosmological constant for the metric (\ref{LineEl.Canonical}), the general
solution being given by (\ref{A1A2}), (\ref{S1S2-2}), and (\ref{Delta1Delta2}%
). Actually, this is, locally, the Kerr-NUT-de Sitter metric, as can be seen
by the change of coordinates $(x,y)\rightarrow(p,q)$%
\begin{equation}
x^{2}=b_{1}^{-1}\left(  p^{2}-\frac{b_{2}}{b_{1}\eta_{2}-b_{2}\eta_{1}%
}\right)  ^{-1}-b_{1}^{-1}\eta_{1}\quad,\quad y^{2}=b_{1}^{-1}\left(
q^{2}+\frac{b_{2}}{b_{1}\eta_{2}-b_{2}\eta_{1}}\right)  ^{-1}+b_{1}^{-1}%
\eta_{1}\,,
\end{equation}
and a relabeling of the integration constants.

\section{Killing-Yano Tensors}

A totally skew-symmetric tensor of rank $p$, $Y_{a_{1}a_{2}\cdots a_{p}%
}=Y_{[a_{1}a_{2}\cdots a_{p}]}$, is called a Killing-Yano (KY) tensor of order
$p$ whenever it obeys the following generalization of the Killing vector
equation:
\begin{equation}
\nabla_{a}\,Y_{b_{1}b_{2}\cdots b_{p}}\,+\,\nabla_{b_{1}}\,Y_{ab_{2}\cdots
b_{p}}\,=\,0\,.
\end{equation}
By means of a KY tensor one can build objects that are conserved along the
geodesic motion. Indeed, if $Y_{a_{1}a_{2}\cdots a_{p}}$ is a Killing-Yano
tensor and $T^{a}$ is an affinely parameterized geodesic vector field,
$T^{a}\nabla_{a}T_{b}=0$, then the tensor $T^{a}Y_{ab_{2}\cdots b_{p}}$ is
constant along each geodesic curve tangent to $\boldsymbol{T}$. As a
consequence, the scalar $Y_{a}^{\phantom{a}c_{2}\cdots c_{p}}Y_{c_{2}\cdots
c_{p}b}T^{a}T^{b}$ is also conserved along the geodesics tangent to
$\boldsymbol{T}$. This, in turn, means that the symmetric tensor
\begin{equation}
Q_{ab}\,=\,Y_{a}^{\phantom{a}c_{2}\cdots c_{p}}\,Y_{c_{2}\cdots c_{p}b}%
\end{equation}
is a Killing tensor of order two. Thus, to each KY tensor it is associated a
Killing tensor of order two, although the converse generally is not true, as
we shall see. Because of this, one can say that KY tensors are, in a sense,
more fundamental than Killing tensors. Physically, this is corroborated by the
fact that classical symmetries associated to KY tensors are preserved at the
quantum level, whereas those associated to Killing tensors generally are not
\cite{Santillan}. In this section we shall investigate whether the Killing
tensor of our metric (\ref{LineEl.Canonical}) is the square of a Killing-Yano
tensor. For a detailed discussion of KY tensors in 4-dimensional spacetimes
the reader is referred to \cite{Dietz,Hall-KY}.

Since $\partial_{\tau}$ and $\partial_{\sigma}$ are Killing vector fields and
the metric is covariantly constant, it follows that the most general Killing
tensor of order two in a manifold with line element (\ref{LineEl.P}) is given
by
\begin{equation}
\boldsymbol{Q}\,=\,\alpha\,\boldsymbol{K}\,+\,\beta\,\boldsymbol{g}%
\,+\,\gamma^{ij}\,\partial_{i}\odot\partial_{j}\,,
\end{equation}
where $\boldsymbol{K}$ is given by (\ref{KillingT1}), $\boldsymbol{g}$ is the
metric tensor and the coefficients $\alpha$, $\beta$ and $\gamma^{ij}$ are
arbitrary constants. Now, for simplicity, let us neglect the terms of
$\boldsymbol{Q}$ coming from the symmetrized products of Killing vectors,
\textit{i.e.}, set $\gamma^{ij}=0$. Then, using (\ref{g-nullTetrad}) along
with (\ref{KillingT2}) lead us to:
\begin{equation}
\boldsymbol{Q}\,=\,\alpha\,\boldsymbol{K}\,+\,\beta\,\boldsymbol{g}%
\,=\,-\,\left(  \,\alpha\,S_{1}\,+\,\beta\,\right)  \,\boldsymbol{l}%
\odot\boldsymbol{n}\,-\,\left(  \,\alpha\,S_{2}\,-\,\beta\,\right)
\,\boldsymbol{m}_{1}\odot\boldsymbol{m}_{2}\,. \label{Q}%
\end{equation}
The goal of the present section is to look for the existence of a Killing-Yano
tensor whose square have the form of the Killing tensor (\ref{Q}). In this
section we shall work with the general metric (\ref{LineEl.Canonical}) without
restricting the functions $A_{1},\,A_{2},\,\Delta_{1},\,\Delta_{2},\,S_{1}$
and $S_{2}$. In particular, Einstein's equation will not be assumed to hold.

Since a KY tensor of order one is just a Killing vector and, in four
dimensions, a KY tensor of order four is a constant multiple of the
volume-form, it follows that the only non-trivial Killing-Yano tensors are the
ones of order two and three. Let us first consider the possibility of
$\boldsymbol{Q}$ being the square of a KY tensor of order three. In this case,
$\boldsymbol{Q}$ would have the following form \cite{Bat-KYn-1}:
\begin{equation}
\label{Q'}Q_{ab} \,=\, \xi_{a}\,\xi_{b} \,-\, (\xi^{c}\xi_{c}) \,g_{ab}\,,
\end{equation}
where $\xi_{a}$ is a conformal Killing vector. However, expanding the vector
$\boldsymbol{\xi}$ in the null tetrad basis and then inserting into (\ref{Q'})
one can easily see that (\ref{Q'}) is equal to (\ref{Q}) only in the trivial
case in which $\xi_{a}=0$ and $\alpha=0=\beta$. Thus, a non-zero
$\boldsymbol{Q}$ cannot be the square of a KY tensor of order three. It
remains to check whether Killing tensor $\boldsymbol{Q}$ in (\ref{Q}) is the
square of a KY tensor of order two.

If a bivector $Y_{ab}$ is such that its square has the algebraic form of
$\boldsymbol{Q}$ in Eq. (\ref{Q}) then it might have the following form:
\begin{equation}
\boldsymbol{Y}\,=\,-\,\Phi_{1}\,\boldsymbol{l}\wedge\boldsymbol{n}%
\,+\,i\,\Phi_{2}\,\boldsymbol{m}_{1}\wedge\boldsymbol{m}_{2} \label{Y-general}%
\end{equation}
where
\begin{equation}
\left(  \,\Phi_{1}\,\right)  ^{2}\,=\,\alpha\,S_{1}\,+\,\beta\quad\text{ and
}\quad\left(  \,\Phi_{2}\,\right)  ^{2}\,=\,\alpha\,S_{2}\,-\,\beta\,.
\label{Phi}%
\end{equation}
Since the integrability condition for the existence of a Killing-Yano tensor
of order two implies that the Petrov type of the Weyl tensor is $D$, $N$ or
$O$, we can attain ourselves to these cases. Nevertheless, since the Weyl
scalars of the metric (\ref{LineEl.P}) are such that $\Psi_{0}=\Psi_{4}=0$ it
follows that the type $N$ is forbidden. Then, using the fact that the type $O$
can be seen as a special case of the type $D$, we conclude that a necessary
condition for $\boldsymbol{Y}$ to be a KY tensor is that the Weyl tensor
should be at least type $D$. Therefore, without loss of generality, we can
assume (\ref{A1A2}) to hold whenever the space with metric
(\ref{LineEl.Canonical}) admits a KY tensor. Then, by means of integrating the
Killing-Yano equation for the bivector (\ref{Y-general}) in a space with the
general metric (\ref{LineEl.Canonical}) along with (\ref{A1A2}), we see that:
besides $\Phi_{1}$ and $\Phi_{2}$, we have that the functions $S_{1}$ and
$S_{2}$ appearing in the metric are also constrained, which can be grasped
from the relation (\ref{Phi}). The final result is that $\boldsymbol{Y}$ is a
KY tensor if, and only if, the functions $\Phi_{1}$, $\Phi_{2}$, $S_{1}$ and
$S_{2}$ are given by:
\begin{align}
\Phi_{1}(x)\,=\,c\,\sqrt{\frac{b_{2}\,x^{2}\,+\,\eta_{2}}{b_{1}\,x^{2}%
\,+\,\eta_{1}}}\quad\;  &  ,\;\quad\Phi_{2}(y)\,=\,c\,\sqrt{\frac
{-\,b_{2}\,y^{2}\,+\,\eta_{2}}{b_{1}\,y^{2}\,-\,\eta_{1}}}\label{Phi-KY}\\
S_{1}(x)\,=\,\frac{b_{3}\,x^{2}+\eta_{3}}{b_{1}\,x^{2}+\eta_{1}}\quad\;  &
,\;\quad S_{2}(y)\,=\,-\,\frac{b_{3}\,y^{2}-\eta_{3}}{b_{1}\,y^{2}-\eta_{1}%
}\,. \label{S-KY}%
\end{align}
Where $c$, $b_{3}$ and $\eta_{3}$ are arbitrary constants,\footnote{Besides
the solution displayed in Eqs. (\ref{Phi-KY}) and (\ref{S-KY}), one also have
a solution if the replacements $b_{1}\leftrightarrow b_{2}$ and $\eta
_{1}\leftrightarrow\eta_{2}$ are performed in (\ref{Phi-KY}) and (\ref{S-KY}).
However, since this other solution can be obtained from the previous one just
by a redefinition of constants that are not fixed yet, we shall consider that
they represent the same solution.} whereas the constants $b_{1}$, $b_{2}$,
$\eta_{1}$ and $\eta_{2}$ are the ones appearing in (\ref{A1A2}). It is
interesting noting that the functions $S_{1}$ and $S_{2}$ compatible with the
existence of a KY tensor of order two are exactly equal to the ones found
while solving Einstein's vacuum equation, see (\ref{S1S2-2}). In particular,
this means that the requirement of the existence of a Killing-Yano tensor in a
space with line element (\ref{LineEl.Canonical}) implies that the eight
components (\ref{Einstein8}) of Einstein's vacuum equation are satisfied. This
hints that often the geometrical requirement of the existence of a KY tensor
might be quite helpful in integrating Einstein's vacuum equation.
Particularly, (\ref{S-KY}) implies that all the vacuum solutions found in Sec.
\ref{Sec.Einstein-General} are endowed with a Killing-Yano tensor. Indeed, it
is well-known that all type $D$ Ricci-flat spacetimes possessing a non-trivial
Killing tensor also have a KY tensor \cite{Collinson-Stephani}. The results of
this section illuminates the possibility that the latter fact can be extended
from Ricci-flat to Einstein spacetimes, namely to the case of non-zero
cosmological constant.

The square of the Killing-Yano tensor $\boldsymbol{Y}$, $Q_{ab}=Y_{a}%
^{\phantom{a}c}Y_{cb}$, is given by
\begin{equation}
\boldsymbol{Q}\,=\,-\,(\Phi_{1})^{2}\,\boldsymbol{l}\odot\boldsymbol{n}%
\,-\,(\Phi_{2})^{2}\,\boldsymbol{m}_{1}\odot\boldsymbol{m}_{2}\,. \label{YY}%
\end{equation}
Comparing (\ref{KillingT2}) and (\ref{YY}) we conclude that in order to have
$\boldsymbol{K}=\boldsymbol{Q}$ the relations $(\Phi_{1})^{2}=S_{1}$ and
$(\Phi_{2})^{2}=S_{2}$ must both hold. However, in general we cannot manage to
choose the constant $c$ appearing in (\ref{Phi-KY}) to be such that the latter
conditions are satisfied. Which lead us to the conclusion that generally there
is no KY tensor whose square is the Killing tensor $\boldsymbol{K}$. Instead,
the square of the KY tensor $\boldsymbol{Y}$ is a linear combination of
$\boldsymbol{K}$ and $\boldsymbol{g}$, as anticipated in Eq. (\ref{Q}).

Note that the functions $\Delta_{1}$ and $\Delta_{2}$ are not constrained by
the Killing-Yano equation. So, there are non-vacuum type $D$ spacetimes that
admit the existence of a KY tensor, which is already clear in Carter's metric
\cite{Carter-KleinG,FrolovBook}. What maybe is not so clear in the literature
and is clarified by our results is that there are type $D$ spacetimes
admitting a non-trivial Killing tensor that do not admit KY tensors. Indeed,
if $A_{1}$ and $A_{2}$ are given by (\ref{A1A2}) and the functions $S_{1}$ and
$S_{2}$ are not of the form displayed in (\ref{S-KY}) then the spacetime with
metric (\ref{LineEl.Canonical}) is type $D$, posses a non-trivial Killing
tensor but does not admit a KY tensor.



\section{Integrating Einstein's Equation when $P_{1}^{\prime}\neq0$ and
$P_{2}^{\prime}= 0$}

In the previous sections we considered the metric (\ref{LineEl.P}) in the
general case when both functions $P_{1}$ and $P_{2}$ are non-constant. Now, we
shall investigate the cases in which at least one of these functions are
constant. Particularly, the aim of the present section is to study the case of
$P_{1}^{\prime}\neq0$ and $P_{2}^{\prime}=0$. More precisely, we shall fully
integrate Einstein's vacuum equation with a cosmological constant and look for
the existence of Killing-Yano tensors in these solutions. Note that it is
needless to consider the analogous case $P_{1}^{\prime}=0$ and $P_{2}^{\prime
}\neq0$, inasmuch as such a case can be easily obtained from the case
$P_{1}^{\prime}\neq0$ and $P_{2}^{\prime}=0$ by interchanging the coordinates
$x$ and $y$.

Since in this section it will be assumed that $P_{1}(x)$ is non-constant, it
follows that we can redefine the coordinate $x$, along with the functions
$A_{1}$ and $\Delta_{1}$, in such a way that $P_{1}(x)=\frac{1}{x}-p_{2}$,
with $p_{2}$ denoting the constant value of the function $P_{2}(y)$. Then,
using this gauge choice, the line element (\ref{LineEl.P}) becomes
\begin{equation}
\label{LineEl.P2-cte}ds^{2}\,=\,S\,\left[  -\,x^{2}\,A_{2}\,\Delta
_{2}\,\left(  \,d\tau+\left(  x^{-1}-p_{2}\right)  \,d\sigma\,\right)
^{2}\,+\,A_{1}\,\Delta_{1}\,x^{2}\left(  \,d\tau-p_{2}\,d\sigma\,\right)
^{2}\,+\,\frac{dx^{2}}{\Delta_{1}}\,+\,\frac{dy^{2}}{\Delta_{2}}\right]  \,
\end{equation}
where $S=S_{1}(x)+S_{2}(y)$. Furthermore, using the coordinate $\phi=\frac
{1}{\ell}(\tau-p_{2}\sigma)$ instead of $\tau$, the line element assumes the
following form:
\begin{equation}
\label{LineEl.P2-cte2}ds^{2}\,=\,S\,\left[  -\,A_{2}\,\Delta_{2}\,\left(
\,d\sigma+x\,\ell\,d\phi\,\right)  ^{2}\,+\,x^{2}\,A_{1}\,\Delta_{1}\,\ell
^{2}\,d\phi^{2}\,+\,\frac{dx^{2}}{\Delta_{1}}\,+\,\frac{dy^{2}}{\Delta_{2}%
}\right]  \,,
\end{equation}
where $\ell$ is a non-zero constant introduced for future convenience. Now,
let us integrate Einstein's vacuum equation for the above line element.

As explained in Sec. \ref{Sec.General}, a necessary condition for the above
metric to be a solution of Einstein's vacuum equation is that the Weyl tensor
might have Petrov type $D$. According to (\ref{A1-P2CteD}), the type $D$
condition holds if, and only if, $A_{1}$ takes the following form
\begin{equation}
A_{1}(x)\,=\,\frac{1}{\,\ell^{2}\,x^{2}}\,. \label{A1-P2Cte-2}%
\end{equation}
Inserting (\ref{A1-P2Cte-2}) into (\ref{LineEl.P2-cte2}) and then computing
the Ricci tensor we find that:
\begin{equation}
R_{\phantom{x}y}^{x}\,=\,\frac{\,3\,\Delta_{1}\,S_{1}^{\prime}\,S_{2}^{\prime
}}{2\,(S_{1}\,+\,S_{2})^{3}\,}\,, \label{Rxy-P2cte}%
\end{equation}
where, as usual, the primes denote that the function is being differentiated
with respect to its variable. Then, imposing $R_{\phantom{a}b}^{a}%
=\Lambda\,\delta_{\phantom{a}b}^{a}$ we conclude that the right hand side of
(\ref{Rxy-P2cte}) must vanish. Thus, either $S_{1}(x)$ or $S_{2}(y)$ might be
constant. In principle, one could also have that both $S_{1}(x)$ or $S_{2}(y)$
are constant. However, assuming $S_{2}(y)$ to be constant we find that
$R_{ab}l^{a}l^{b}$ does not vanish as it should. Therefore, we conclude that
$S_{2}(y)$ should be a non-constant function, while $S_{1}(x)$ is a constant
that we shall denote by $s_{1}$. Thus, the conformal factor $S(x,y)$ should be
just a function of $y$:
\begin{equation}
S(x,y)\,=\,s_{1}\,+\,S_{2}(y)\,=\,S(y)\,.
\end{equation}
Now, without loss of generality, let us choose the coordinate $y$ in such a
way that
\begin{equation}
S(y)=y^{2}+n_{1}^{2}\, \label{S(y)}%
\end{equation}
with $n_{1}$ being a constant. Since the value of $n_{1}$ can be shifted by
means of redefining the coordinate $y$, in what follows it will be assumed
that $n_{1}\neq0$. Then, assuming (\ref{S(y)}) and imposing $R_{ab}l^{a}l^{b}$
to vanish it follows that $A_{2}(y)$ must be given by:
\begin{equation}
A_{2}(y)\,=\,\frac{4y^{2}}{\,\left(  4\,n_{1}^{2}\,n_{2}^{2}-\ell^{2}%
+4\,n_{2}^{2}\,y^{2}\right)  \left(  n_{1}^{2}+y^{2}\right)  ^{2}}\,,
\label{A2-P2-cte0}%
\end{equation}
with $n_{2}$ being an integration constant. Postponing the special case
$n_{2}=0$ to the forthcoming section, let us assume $n_{2}\neq0$. In the
latter case we can choose the non-zero parameter $\ell$ to be equal to
$2n_{1}n_{2}$, in which case we have
\begin{equation}
A_{2}(y)\,=\,\frac{1}{\,n_{2}^{2}\left(  n_{1}^{2}+y^{2}\right)  ^{2}}\,.
\label{A2-P2-cte}%
\end{equation}
Once assumed the latter expression for $A_{2}$, the eight components of
Einstein's vacuum equation displayed in (\ref{Einstein8}) are immediately
satisfied. Finally, imposing the equation $R_{ab}\,m_{1}^{\,a}m_{2}%
^{\,b}=\Lambda$ we find that the functions $\Delta_{1}$ and $\Delta_{2}$
should have the following general form:
\begin{equation}
\Delta_{1}(x)\,=-\,a_{2}\,x^{2}\,+\,a_{1}\,x\,+\,a_{0}\,, \label{Delta1-P2cte}%
\end{equation}%
\begin{equation}
\Delta_{2}(y)\,=\left(  n_{1}^{4}-2\,n_{1}^{2}\,y^{2}-\frac{1}{3}%
\,y^{4}\right)  \Lambda\,+\,a_{2}\,(y^{2}-n_{1}^{2})+b\,y\,,
\label{Delta2-P2-Cte}%
\end{equation}
with $a_{0}$, $a_{1}$, $a_{2}$ and $b$ being integration constants. In
particular, $b$ is related to the ADM mass of the solution. In conclusions,
the general solution of Einstein's vacuum equation with a cosmological
constant for the metric (\ref{LineEl.P}) with $P_{2}$ constant and $P_{1}$
non-constant is given by the equations (\ref{LineEl.P2-cte2}),
(\ref{A1-P2Cte-2}), (\ref{S(y)}), (\ref{A2-P2-cte}), (\ref{Delta1-P2cte}) and
(\ref{Delta2-P2-Cte}).
It turns out that such solution posses four Killing vector fields. Indeed,
defining $\omega=\frac{1}{2}\sqrt{a_{1}^{2}+4a_{0}a_{2}}$, it can be verified
that that the following vector fields generate isometries:
\begin{align}
\boldsymbol{\chi}_{1}\,=  &  \,-\,\sin(\omega\phi)\,\frac{n_{1}\,n_{2}%
\,(2\,a_{0}+a_{1}x)}{\omega\,\sqrt{\Delta_{1}}}\,\partial_{\sigma}%
\,+\,\sin(\omega\phi)\,\frac{2\,a_{2}\,x-a_{1}}{2\,\omega\,\sqrt{\Delta_{1}}%
}\,\partial_{\phi}\,+\,\cos(\omega\phi)\,\sqrt{\Delta_{1}}\,\partial_{x}\\
\boldsymbol{\chi}_{2}\,=  &  \,\,\cos(\omega\phi)\,\frac{n_{1}\,n_{2}%
\,(2\,a_{0}+a_{1}x)}{\omega\,\sqrt{\Delta_{1}}}\,\partial_{\sigma}%
\,-\,\cos(\omega\phi)\,\frac{2\,a_{2}\,x-a_{1}}{2\,\omega\,\sqrt{\Delta_{1}}%
}\,\partial_{\phi}\,+\,\sin(\omega\phi)\,\sqrt{\Delta_{1}}\,\partial_{x}\,,
\end{align}
in addition to the obvious Killing vector fields $\boldsymbol{\chi}%
_{3}=\partial_{\phi}$ and $\boldsymbol{\chi}_{4}=\partial_{\sigma}$.


Since $S=S_{1}+S_{2}=s_{1}+S_{2}$, it follows that the we can absorb the
constant $s_{1}$ into the function $S_{2}$ so that instead of using the
functions $S_{1}$ and $S_{2}$ one could equivalently use $\tilde{S}_{1}(x)=0$
and $\tilde{S}_{2}(y)=s_{1}+S_{2}=S$. Thus, besides the Killing tensor
(\ref{KillingT2}), we expect that the tensor
\begin{equation}
\boldsymbol{K}_{2}\,=\,-\,\tilde{S}_{1}(x)\,\boldsymbol{l}\odot\boldsymbol{n}%
\,-\,\tilde{S}_{2}(y)\,\boldsymbol{m}_{1}\odot\boldsymbol{m}_{2}%
\,=\,\,-\,S(y)\,\boldsymbol{m}_{1}\odot\boldsymbol{m}_{2}%
\end{equation}
should also be a Killing tensor. Indeed, this can be readily verified.
However, it turns out that this new Killing tensor does not lead to new
conserved charges, which can be grasped from the fact that $\boldsymbol{K}%
_{2}$ is just a linear combination of $\boldsymbol{K}$ and the metric,
$\boldsymbol{K}_{2}=\boldsymbol{K}-s_{1}\boldsymbol{g}$. Moreover, the latter
Killing tensor is reducible, in the sense that it can be written in terms of
symmetrized products of Killing vectors. Indeed, one can check that
\begin{equation}
\boldsymbol{K}_{2}\,=\,\frac{2\,a_{0}\,n_{1}^{2}\,n_{2}^{2}}{\omega^{2}%
}\,\boldsymbol{\chi}_{4}\odot\boldsymbol{\chi}_{4}\,+\,\frac{a_{1}%
\,n_{1}\,n_{2}}{\omega^{2}}\,\boldsymbol{\chi}_{4}\odot\boldsymbol{\chi}%
_{3}\,-\,\frac{a_{2}}{2\,\omega^{2}}\,\boldsymbol{\chi}_{3}\odot
\boldsymbol{\chi}_{3}\,-\,\frac{1}{2}\,\boldsymbol{\chi}_{1}\odot
\boldsymbol{\chi}_{1}\,-\,\,\frac{1}{2}\,\boldsymbol{\chi}_{2}\odot
\boldsymbol{\chi}_{2}\,.
\end{equation}
The solution found in this section also posses a Killing-Yano tensor given by
\begin{equation}
\boldsymbol{Y}\,=\,n_{1}\,\boldsymbol{l}\wedge\boldsymbol{n}%
\,+\,i\,y\,\boldsymbol{m}_{1}\wedge\boldsymbol{m}_{2}\,,
\end{equation}
whose square is $\boldsymbol{K}_{2}+n_{1}^{2}\,\boldsymbol{g}$.

Regarding the interpretation of the latter metric, the existence of four
Killing vectors hints the existence of spherical symmetry and that such
solution might be a generalization of the Schwarzschild metric. Indeed, if
$a_{2}\neq0$, it follows that the killing vector fields
\begin{equation}
\tilde{\boldsymbol{\chi}}_{1} \,=\, \frac{1}{a_{2}}\,\boldsymbol{\chi}_{1}
\;\;, \quad\tilde{\boldsymbol{\chi}}_{2} \,=\, \frac{1}{\sqrt{a_{2}}%
}\,\boldsymbol{\chi}_{2} \;\;, \quad\tilde{\boldsymbol{\chi}}_{3} \,=\,
-\,\frac{1}{\Lambda\,\sqrt{a_{2}}}\,\left(  \boldsymbol{\chi}_{3} -
\frac{a_{1}\,n_{1}\,n_{2}}{a_{2}} \boldsymbol{\chi}_{4} \right)  \,,
\end{equation}
generate the $SO(3)$ Lie algebra
\begin{equation}
[ \tilde{\boldsymbol{\chi}}_{i}, \tilde{\boldsymbol{\chi}}_{j}] \,=\,
\varepsilon_{ij}^{\phantom{ij}k}\,\tilde{\boldsymbol{\chi}}_{k} \,,
\end{equation}
with $\varepsilon_{ij}^{\phantom{ij}k}$ denoting the usual Levi-Civita symbol.
Therefore, in the case $a_{2}\neq0$, the isometry group is $SO(3)\times
\mathbb{R}$, with $\boldsymbol{\chi}_{4}$ spanning the center of the algebra.
This gives a clue that the Taub-NUT solution with a cosmological constant
might be contained in the class of metrics that we have just found. Indeed,
assuming $a_{1}=0$ and defining new coordinates $\{t,y,\theta,\varphi\}$ by
$\sigma= n_{2} t$, $x=\sqrt{\frac{a_{0}}{a_{2}}}\cos\theta$ and $\varphi
=-\sqrt{a_{0}a_{2}}\phi$ we have that the metric can be written as
\begin{equation}
\label{LineEl.P2-cte3}ds^{2}\,=\,-\, \frac{\Delta_{2}}{y^{2} \,+\, n_{1}^{2}%
}\,\left(  \,dt-\frac{2\,n_{1}}{a_{2}}\,\cos\theta\,\,d\varphi\,\right)  ^{2}
\,+\, \frac{y^{2} \,+\, n_{1}^{2}}{\Delta_{2}}\, dy^{2} \,+\, \frac{y^{2}
\,+\, n_{1}^{2}}{a_{2}}\, \left(  \, d\theta^{2} \,+\, \sin^{2}\theta
\,\,d\varphi^{2} \,\right)  \,,
\end{equation}
with $\Delta_{2}(y)$ given by (\ref{Delta2-P2-Cte}). Making $a_{2}=1$ and
$\Lambda=0$ we get the Taub-NUT solution in the form presented in
\cite{GrifPodol-Book}, with $n_{1}$ being the NUT parameter and $-b/2$ being
the mass. On the other hand, in the special case in which $a_{2} = 0$, the
isometry Lie algebra is not the direct sum of an abelian algebra and a
semi-simple Lie algebra. Indeed, in such a case we have that
\begin{equation}
[ \boldsymbol{\chi}_{1}, \boldsymbol{\chi}_{3}] \,=\,\frac{a_{1}}{2}\,
\boldsymbol{\chi}_{2} \quad,\quad[ \boldsymbol{\chi}_{2}, \boldsymbol{\chi
}_{3}] \,=\, -\,\frac{a_{1}}{2}\, \boldsymbol{\chi}_{1} \quad,\quad[
\boldsymbol{\chi}_{1}, \boldsymbol{\chi}_{2}] \,=\, 2\,n_{1} n_{2}\,
\boldsymbol{\chi}_{4} \,,
\end{equation}
with all other commutators being zero.

\subsection{The special case $n_{2}=0$}

Now, let us consider the special case in which the integration constant
$n_{2}$ vanishes. In such a case, Eq. (\ref{A2-P2-cte0}) gives
\begin{equation}
A_{2}(y)=\frac{-\,4y^{2}}{\,\ell^{2}\left(  n_{1}^{2}+y^{2}\right)  ^{2}}\,.
\end{equation}
Assuming the latter expression for $A_{2}$, the eight components of Einstein's
vacuum equation displayed in (\ref{Einstein8}) are immediately satisfied.
Finally, imposing the equation $R_{ab}\,m_{1}^{\,a}m_{2}^{\,b}=\Lambda$ we
find that the functions $\Delta_{1}$ and $\Delta_{2}$ might have the following
general form:
\begin{equation}
\Delta_{1}(x)\,=-\,a_{2}\,x^{2}\,+\,a_{1}\,x\,+\,a_{0} \label{D1}%
\end{equation}%
\begin{equation}
\Delta_{2}(y)\,=\frac{b}{y^{2}}\,-\,\frac{\Lambda}{2}\left(  n_{1}^{4}%
+n_{1}^{2}\,y^{2}+\frac{1}{3}\,y^{4}\right)  +\frac{a_{2}}{4}\,\left(
2\,n_{1}^{2}+y^{2}\right)  \,, \label{Delta2-P2-Cte2}%
\end{equation}
One can check that the functions $A_{2}$ and $\Delta_{2}$ can be conveniently
written in terms of the function $S(y)$ as follows:
\begin{equation}
A_{2}(y)=-\,\left(  \,\frac{S^{\prime}}{\ell\,S}\,\right)  ^{2}\quad
\;,\;\quad\Delta_{2}(y)\,=\,\frac{1}{\left(  \,S^{\prime}\,\right)  ^{2}%
}\,\left[  \,\tilde{b}\,+\,a_{2}\,S^{2}\,-\,\frac{2}{3}\,\Lambda
\,S^{3}\,\right]  \,. \label{A2D2}%
\end{equation}
Where $\tilde{b}\equiv(4b-a_{2}\,n_{1}^{4}+\frac{2}{3}\Lambda n_{1}^{6})$ is a
constant that replaces the arbitrary constant $b$. It is worth pointing out
that the functions $A_{2}$ and $\Delta_{2}$ as written in (\ref{A2D2}) provide
a solution for Einstein's vacuum equation irrespective of the choice of
coordinate $y$. Thus, if we use (\ref{A2D2}) is not necessary to assume that
$S(y)$ is given by (\ref{S(y)}). As we shall see in the sequel, it turns out
that the metric given by (\ref{LineEl.P2-cte2}), (\ref{A1-P2Cte-2}),
(\ref{D1}), and (\ref{A2D2}) is quite special, since it admits a covariantly
constant bivector whose square is the metric.

But, before proceeding, note that since $A_{1}$ and $A_{2}$ have opposite
signs it follows that this metric cannot have Lorentzian signature. Indeed, by
means of studying the reality conditions \cite{Bat1} of the null tetrad
(\ref{NullTetrad1}), one can see that: If $\ell^{2}<0$ the signature is split
(neutral), while if $\ell^{2}>0$ we have that the signature is Euclidian for
$\Delta_{1}\Delta_{2}>0$ and split for $\Delta_{1}\Delta_{2}<0$. Furthermore,
the bivectors $\boldsymbol{l}\wedge\boldsymbol{n}$ and $\boldsymbol{m}%
_{1}\wedge\boldsymbol{m}_{2}$ are both real if $\ell^{2}<0$ and both imaginary
if $\ell^{2}>0$. Therefore, it is useful to separate our analysis in two cases.

Let us start considering the case $\ell^{2}>0$. In this case we have that the
following real bivector is covariantly constant:
\begin{equation}
\boldsymbol{\Omega}\,=\,-\,i\,\left(  \,\boldsymbol{l}\wedge\boldsymbol{n}%
\,+\,\epsilon\,\,\boldsymbol{m}_{1}\wedge\boldsymbol{m}_{2}\,\right)  \,.
\end{equation}
Where $\epsilon=\pm1$, depending on the function $S$ and on the patch of the
coordinate $x$. More precisely, we have
\begin{equation}
\epsilon\,=\,\text{Sign}\left[  \,x\,\frac{S^{\prime}}{S}\,\right]
\,=\,\pm\,1\,. \label{eps}%
\end{equation}
The bivector $\boldsymbol{\Omega}$ is anti-self-dual if $\epsilon=1$, namely
its Hodge dual is equal to the negative of itself, whereas if $\epsilon=-1$ it
follows that $\boldsymbol{\Omega}$ is self-dual, \textit{i.e.}, its Hodge dual
is equal to itself.\footnote{It is worth recalling that, locally, the
distinction between self-dual and anti-self-dual forms is just a matter of
convention, since by multiplying the volume-form by $-1$ these labels get
interchanged.} Since we have that $\Omega^{ac}\Omega_{cb}=-\delta_{\,b}^{a}$,
we say that the tensor $\boldsymbol{\Omega}$ is an almost complex structure.
Note that the vectors $\boldsymbol{l}$, $\boldsymbol{m}_{1}$, $\boldsymbol{n}$
and $\boldsymbol{m}_{2}$ are eigenvectors of $\boldsymbol{\Omega}$ with
eigenvalues $\pm i$,
\begin{equation}
\Omega_{\phantom{a}b}^{a}\,l^{b}\,=\,i\,l^{a}\;,\;\Omega_{\phantom{a}b}%
^{a}\,n^{b}\,=\,-\,i\,n^{a}\;,\;\Omega_{\phantom{a}b}^{a}\,m_{1}%
^{b}\,=\,-\,i\,\epsilon\,m_{1}^{a}\;,\;\Omega_{\phantom{a}b}^{a}\,m_{2}%
^{b}\,=\,i\,\epsilon\,m_{2}^{a}\,.
\end{equation}
Moreover, irrespective of the sing of $\epsilon$, the eigenspaces of
$\boldsymbol{\Omega}$ form integrable distributions. Indeed, as a consequence
of (\ref{IntegrableDist}), it follows that the isotropic planes generated by
$\{\boldsymbol{l},\boldsymbol{m}_{1}\}$, $\{\boldsymbol{n},\boldsymbol{m}%
_{2}\}$, $\{\boldsymbol{l},\boldsymbol{m}_{2}\}$ and $\{\boldsymbol{n}%
,\boldsymbol{m}_{1}\}$ are all tangent to integrable foliations. Because of
this, we say that such almost complex structure is integrable
\cite{Bat-Book-art2}. Then, since $\boldsymbol{\Omega}$ is a closed form,
$d\boldsymbol{\Omega}=0$, this 2-form is named a K\"{a}hler form. Thus, the
solution found here is a K\"{a}hler metric. Particularly, if $\Lambda=0$ we
end up with a Ricci-flat K\"{a}hler metric, also known as a Calabi-Yau
manifold. In addition, this space is also endowed with the following real
conformal Killing-Yano tensor
\begin{equation}
\boldsymbol{C}\,=\,\,i\,S(y)\,\left(  \,\boldsymbol{l}\wedge\boldsymbol{n}%
\,-\,\epsilon\,\boldsymbol{m}_{1}\wedge\boldsymbol{m}_{2}\,\right)  \,,
\end{equation}
which is a self-dual bivector if $\epsilon>1$ and anti-self-dual if
$\epsilon<1$.

On the other hand, if $\ell^{2}<0$ we have that the real covariantly constant
bivector is given by
\begin{equation}
\check{\boldsymbol{\Omega}}\,=\,\boldsymbol{l}\wedge\boldsymbol{n}%
\,-\,\epsilon\,\boldsymbol{m}_{1}\wedge\boldsymbol{m}_{2}\,,
\end{equation}
with $\epsilon$ given again by (\ref{eps}). In this case we have that
$\check{\Omega}^{ac}\check{\Omega}_{cb}=\delta_{\,b}^{a}$, so that
$\check{\boldsymbol{\Omega}}$ is called an almost paracomplex structure
\cite{ParaComplex}. Because of the integrability of the eigen-planes of this
paracomplex structure we say that it is integrable. Furthermore, since
$\check{\boldsymbol{\Omega}}$ is a closed form this 2-form is named a
para-K\"{a}hler form, so that the metric represents a para-K\"{a}hler manifold
\cite{ParaComplex}. When $\ell^{2}<0$ we also have that the following real
bivector
\begin{equation}
\check{\boldsymbol{C}}\,=\,\,S(y)\,\left(  \,\boldsymbol{l}\wedge
\boldsymbol{n}\,+\,\epsilon\,\boldsymbol{m}_{1}\wedge\boldsymbol{m}%
_{2}\,\right)
\end{equation}
is a conformal Killing-Yano tensor.


In addition to these geometrical objects, the space described here admits four
null bivectors that are solutions of source-free Maxwell equations
irrespective of the sign of the constant $\ell^{2}$:
\begin{equation}
\boldsymbol{B}_{1}^{+}\,=\,\frac{1}{S^{\prime}\,\sqrt{\Delta_{1}\,\Delta_{2}}%
}\,\boldsymbol{m}_{2}\wedge\boldsymbol{n}\quad,\quad\boldsymbol{B}_{2}%
^{+}\,=\,\frac{1}{S^{\prime}\,\sqrt{\Delta_{1}\,\Delta_{2}}}\,\boldsymbol{l}%
\wedge\boldsymbol{m}_{1}\,,
\end{equation}%
\begin{equation}
\boldsymbol{B}_{1}^{-}\,=\,\frac{1}{S^{\prime}\,\sqrt{\Delta_{1}\,\Delta_{2}}%
}\,\boldsymbol{m}_{1}\wedge\boldsymbol{n}\quad,\quad\boldsymbol{B}_{2}%
^{-}\,=\,\frac{1}{S^{\prime}\,\sqrt{\Delta_{1}\,\Delta_{2}}}\,\boldsymbol{l}%
\wedge\boldsymbol{m}_{2}\,.
\end{equation}
The bivectors $\boldsymbol{B}_{1}^{+}$ and $\boldsymbol{B}_{2}^{+}$ are
self-dual, while $\boldsymbol{B}_{1}^{-}$ and $\boldsymbol{B}_{2}^{-}$ are
anti-self-dual. Since these bivectors are closed and co-closed we say that
they obey the source-free Maxwell equations. Actually, since the
energy-momentum tensor associated to these Maxwell fields is zero, we can say
that they provide solutions to Einstein-Maxwell equations.

In order to find possible singularities of the space it is useful to take a
look at some curvature invariant scalars, \textit{i.e.}, scalars that are
constructed from full contractions of the curvature and its derivatives. Note,
for instance, that the Weyl scalars are not curvature invariants, since they
depend on the choice of the null tetrad basis. However, the following scalars
are true curvature invariants:
\begin{align}
R^{abcd}\,R_{abcd}  &  \,=\, \frac{16}{3}\,\Lambda^{2} \,+\, 24\,\left(
\frac{\tilde{b}}{S^{3}} \right)  ^{2} \,,\label{Curvature_scalar1}\\
R^{abcd}\,R_{cdef}\,R^{ef}_{\phantom{ef}ab}  &  \,=\, \frac{80}{9}%
\,\Lambda^{3} \,+\, 48 \left(  \frac{\tilde{b} }{S^{3}} \right)  ^{2} \,-\,
48\,\left(  \frac{\tilde{b} }{S^{3}} \right)  ^{3} \,,
\label{Curvature_scalar2}%
\end{align}
where $R_{abcd}$ stands for the Riemann tensor. Note that these scalars
diverge in the points in which the function $S(y)$ vanishes, hinting the
existence of singularities in these points. Nevertheless, it is interesting
noting that these divergences cease to exist if the constant $\tilde{b}$
vanishes. With the aim of understanding the meaning of the condition
$\tilde{b}=0$ let us compute the Weyl scalars of this space.

Since the space considered in the present section is type $(D,D)$ according to
the generalized Petrov classification, with $\boldsymbol{l}\wedge
\boldsymbol{m}_{1}$, $\boldsymbol{n}\wedge\boldsymbol{m}_{2}$, $\boldsymbol{l}%
\wedge\boldsymbol{m}_{2}$ and $\boldsymbol{n}\wedge\boldsymbol{m}_{1}$ being
repeated principal null bivectors \cite{Bat-Book-art2}, it follows that the
only Weyl scalars that can be different from zero are $\Psi_{2}^{+}$ and
$\Psi_{2}^{-}$. One can check that their values depend on the sign of the
parameter $\ell^{2}$. Indeed, if $\ell^{2}>0$ we find that
\begin{equation}
\Psi_{2}^{+}\,=\,-\,(1+\epsilon)\,\frac{\Lambda}{6}\,+\,(1-\epsilon
)\,\frac{\tilde{b}}{2\,S^{3}}\quad\text{ and }\quad\Psi_{2}^{-}%
\,=\,-\,(1-\epsilon)\,\frac{\Lambda}{6}\,+\,(1+\epsilon)\,\frac{\tilde{b}%
}{2\,S^{3}}\,.
\end{equation}
So, if $\tilde{b}=0$ and $\epsilon=1$ the space is self-dual, meaning that
only the self-dual part of the Weyl tensor is different from zero, whereas if
$\tilde{b}=0$ and $\epsilon=-1$ the space is anti-self-dual. Analogously, if
$\Lambda=0$ the space is anti-self-dual for $\epsilon=1$ and self-dual for
$\epsilon=-1$. On the other hand, if $\ell^{2}<0$ the values of $\Psi_{2}^{+}$
and $\Psi_{2}^{-}$ are interchanged. More explicitly, if $\ell^{2}$ is
negative we have that:
\begin{equation}
\Psi_{2}^{+}\,=\,-\,(1-\epsilon)\,\frac{\Lambda}{6}\,+\,(1+\epsilon
)\,\frac{\tilde{b}}{2\,S^{3}}\quad\text{ and }\quad\Psi_{2}^{-}%
\,=\,-\,(1+\epsilon)\,\frac{\Lambda}{6}\,+\,(1-\epsilon)\,\frac{\tilde{b}%
}{2\,S^{3}}\,.
\end{equation}
Thus, when $\ell^{2}<0$ and $\epsilon=1$ the space is self-dual if $\Lambda=0$
and anti-self-dual if $\tilde{b}=0$. Analogously, if $\ell^{2}<0$ and
$\epsilon=-1$ the space is anti-self-dual if $\Lambda=0$ and self-dual if
$\tilde{b}=0$. So, we conclude the condition $\tilde{b}=0$ that avoids the
divergence of the curvature invariants (\ref{Curvature_scalar1}) and
(\ref{Curvature_scalar2}) means geometrically that the Weyl tensor is either
self-dual or anti-self-dual.


\section{ Integrating Einstein's Equation when $P_{1}^{\prime}=0$ and
$P_{2}^{\prime}= 0$}

The aim of the present section is to integrate Einstein's vacuum equation for
the metric (\ref{LineEl.P}) in the special case in which the functions
$P_{1}(x)$ and $P_{2}(y)$ are both constant. In what follows we shall denote
these constants by $p_{1}$ and $p_{2}$ respectively. In this case, we can
redefine the coordinates $x$ and $y$ along with the functions $\Delta_{1}$,
$\Delta_{2}$, $A_{1}$ and $A_{2}$ in such a way to make $\Delta_{1}(x)=1$ and
$\Delta_{2}(y)=1$. Adopting these redefined coordinates we end up with the
following line element:
\begin{equation}
ds^{2}\,=\,S\,\left[  \frac{-\,A_{2}}{(p_{1}+p_{2})^{2}}\,\left(
\,d\tau+p_{1}\,d\sigma\,\right)  ^{2}\,+\,\frac{A_{1}}{(p_{1}+p_{2})^{2}%
}\,\left(  \,d\tau-p_{2}\,d\sigma\,\right)  ^{2}\,+\,dx^{2}\,+\,dy^{2}\right]
\, \label{LineEl.P1P2-cte}%
\end{equation}
where $S=S_{1}(x)+S_{2}(y)$. As anticipated in Sec. \ref{Sec.General}, this
metric is type $D$ regardless of any restriction on the functions $A_{1}$ and
$A_{2}$. Now, computing the Ricci tensor we find that $R_{\phantom{x}y}^{x}$
is given by the expression (\ref{Rxy-P2cte}) with $\Delta_{1}(x)=1$.
Therefore, in order for Einstein's vacuum equation with a cosmological
constant to be satisfied, either $S_{1}(x)$ or $S_{2}(y)$ might be constant.
One could also have that both functions are constant, but let us postpone the
analysis of this case. So, let us assume that $S_{1}(x)$ is a constant denoted
by $s_{1}$ and that $S_{2}(y)$ is a non-constant function of $y$.\footnote{The
opposite case, in which $S_{2}$ is constant and $S_{1}$ is non-constant can be
obtained from the case $S_{1}^{\prime}=0$ and $S_{2}^{\prime}\neq0$ by means
of interchanging the coordinates $x$ and $y$.} For future convenience, let us
define the function $H(y)$:
\begin{equation}
H(y)\,=\,S^{1/4}\,=\,\left[  \,s_{1}\,+\,S_{2}(y)\,\right]  ^{1/4}\,.
\label{S-h}%
\end{equation}
Then, imposing $R_{ab}\,l^{a}l^{b}$ to vanish and assuming $H(y)$ to be
non-constant, we find that $A_{2}(y)$ must have the following general form:
\begin{equation}
A_{2}(y)\,=\,a_{2}\,\frac{\left(  \,H^{\prime}\,\right)  ^{2}}{H^{6}}\,,
\label{A2-P1P2cte}%
\end{equation}
with $a_{2}$ being an arbitrary non-zero constant. Assuming (\ref{A2-P1P2cte}%
), we have that the eight components of Einstein's vacuum equation displayed
in (\ref{Einstein8}) are satisfied. It remains to impose $R_{ab}\,m_{1}%
^{\,a}m_{2}^{\,b}=\Lambda$ and $R_{ab}\,l^{a}n^{b}=-\Lambda$. The first of
these conditions imply that $A_{1}$ is given by
\begin{equation}
A_{1}(x)\,=\,a_{1}\,\,\cos^{2}(2\,b\,x\,+\,c)\,, \label{A1-P1P2cte}%
\end{equation}
where $a_{1}$, $b$ and $c$ are constants. Inserting this expression for
$A_{1}$ into $R_{ab}\,m_{1}^{\,a}m_{2}^{\,b}=\Lambda$ yields the following
differential equation for $H$:
\begin{equation}
H^{\prime\prime}\,=\,H\,\left(  \,b^{2}\,-\,\frac{\Lambda}{4}\,H^{4}\,\right)
\,. \label{H-DiffEq}%
\end{equation}
One can also prove that if (\ref{H-DiffEq}) holds then the remaining equations
$R_{ab}\,m_{1}^{\,a}m_{2}^{\,b}=\Lambda$ and $R_{ab}\,l^{a}n^{b}=-\Lambda$ are
both obeyed. Particularly, if $\Lambda=0$ the general solution of
(\ref{H-DiffEq}) is given by
\begin{equation}
H(y)\,=\,a_{3}\,e^{b\,y}\,+\,a_{4}\,e^{-\,b\,y}\quad\;\;\left(  \,\Lambda
\,=\,0\,\right)  \,,
\end{equation}
where $a_{3}$ and $a_{4}$ are arbitrary constants. It is worth noting that if
either $a_{3}$ or $a_{4}$ vanish then $\Psi_{2}=0$ and the space is flat. For
$\Lambda\neq0$, any non-constant solution for the non-linear differential
equation (\ref{H-DiffEq}) will generate a metric that is solution of
Einstein's vacuum equation. This solution turns out to admit the following
Killing-Yano tensor:
\begin{equation}
\boldsymbol{Y}\,=\,i\,H^{2}\,\boldsymbol{m}_{1}\wedge\boldsymbol{m}_{2}\,.
\label{KY-P1P2cte1}%
\end{equation}

Note that we can easily get rid of some constants in the solution
(\ref{LineEl.P1P2-cte}) by means of redefining the coordinates as follows:
\begin{equation}
\tilde{\tau}\,=\,\frac{\sqrt{a_{2}}\left(  \,\tau\,+\,p_{1}\,\sigma\,\right)
}{p_{1}\,+\,p_{2}}\quad,\quad\tilde{\sigma}\,=\,2\,\frac{\sqrt{a_{1}}\left(
\,\tau\,-\,p_{2}\,\sigma\,\right)  }{p_{1}\,+\,p_{2}}\quad,\quad\tilde
{x}\,=\,2\,x\,+\,\frac{c}{b}\,+\,\frac{\pi}{2\,b}\quad,\quad\tilde
{y}\,=\,\frac{1}{2}\,\left[  \,H(y)\,\right]  ^{2}\,.
\end{equation}
With these coordinates, the solution just obtained is given by
\begin{equation}
ds^{2}\,=\,\,-\,\left(  \frac{H^{\prime}}{H}\right)  ^{2}\,d\tilde{\tau}%
^{2}\,+\,\left(  \frac{H^{\prime}}{H}\right)  ^{-2}d\tilde{y}^{2}%
\,+\,\tilde{y}^{2}\,\left(  \,d\tilde{x}^{2}\,+\,\sin^{2}(b\,\tilde
{x})\,d\tilde{\sigma}^{2}\,\right)  \,\,. \label{LineEl.P1P2-newCoord1}%
\end{equation}
With $H(y)$ being a non-constant solution of (\ref{H-DiffEq}). Although
(\ref{H-DiffEq}) is a non-linear differential equation, we can transform this
equation into a linear equation by means of using the coordinate $\tilde{y}$.
Indeed, defining $F(\tilde{y})\equiv\left(  \frac{H^{\prime}}{H}\right)  ^{2}$
we find that (\ref{H-DiffEq}) is equivalent to the differential equation
\begin{equation}
\tilde{y}\,\,\frac{dF}{d\tilde{y}}\,+\,F\,=\,b^{2}\,-\,\Lambda\,\tilde{y}%
^{2}\,,
\end{equation}
whose general solution is
\begin{equation}
\left(  \frac{H^{\prime}}{H}\right)  ^{2}\,=\,F(\tilde{y})\,=\,b^{2}%
\,-\,\frac{2\,m}{\tilde{y}}\,-\,\frac{\Lambda}{3}\,\tilde{y}^{2}\,.
\end{equation}
Where $m$ is an integration constant. Therefore, the solution given by
(\ref{LineEl.P1P2-newCoord1}) along with (\ref{H-DiffEq}) is just the
Schwarzschild-(A)dS spacetime with a possible conical singularity. In terms of
these coordinates the null tetrad (\ref{NullTetrad1}) is given by:
\begin{align}
\boldsymbol{l}  &  =-\,\frac{1}{\sqrt{2\,F}}\,\left(  \,F\,d\tilde{\tau
}\,-\,d\tilde{y}\,\right)  \,,\\
\boldsymbol{n}  &  =-\,\frac{1}{\sqrt{2\,F}}\,\left(  \,F\,d\tilde{\tau
}\,+\,d\tilde{y}\,\right)  \,,\\
\boldsymbol{m}_{1}  &  =-\,\frac{\tilde{y}}{\sqrt{2}}\,\left(  \,\sin
(b\,\tilde{x})\,d\tilde{\sigma}\,-\,i\,d\tilde{x}\,\right)
\,,\label{NullTetrad2}\\
\boldsymbol{m}_{2}  &  =-\,\frac{\tilde{y}}{\sqrt{2}}\,\left(  \,\sin
(b\,\tilde{x})\,d\tilde{\sigma}\,+\,i\,d\tilde{x}\,\right)  \,.
\end{align}
In particular, by means of (\ref{KY-P1P2cte1}) and (\ref{NullTetrad2}), we
arrive at the following expression for the Killing-Yano tensor in these new
coordinates:
\begin{equation}
\boldsymbol{Y}\,=\,2\,\tilde{y}^{3}\,\sin(b\,\tilde{x})\,d\tilde{x}\wedge
d\tilde{\sigma}\,.
\end{equation}


\subsection{The case $S_{1}$ and $S_{2}$ constant}

In order to obtain (\ref{A2-P1P2cte}) it was assumed that $S_{2}(y)$ is
non-constant. Now, it is time to consider the case when the functions $P_{1}$,
$P_{2}$, $S_{1}$ and $S_{2}$ are all constant, in which case the component
$R_{ab}\,l^{a}l^{b}$ is automatically zero and there is no constraint over
$A_{2}(y)$ at this stage. Actually, one can check that the eight components
(\ref{Einstein8}) of Einstein's vacuum equation are already satisfied. Then,
the remaining equations $R_{ab}\,m_{1}^{\,a}m_{2}^{\,b}=\Lambda$ and
$R_{ab}\,l^{a}n^{b}=-\Lambda$ provide non-linear differential equations for
$A_{1}$ and $A_{2}$ respectively whose general solutions are:
\begin{equation}
A_{1}(x)\,=\,a_{1}\,\cos^{2}(x\,\sqrt{s\,\Lambda}\,+\,b_{1})\quad\text{ and
}\quad A_{2}(y)\,=\,a_{2}\,\cos^{2}(y\,\sqrt{s\,\Lambda}\,+\,b_{2})\,,
\label{A1A2-P1P2cte}%
\end{equation}
where the $a$'s, the $b$'s and $s\equiv S$ are constants. The above solution
is valid only for $\Lambda\neq0$. Instead, if $\Lambda=0$ the equations
$R_{ab}\,m_{1}^{\,a}m_{2}^{\,b}=\Lambda$ and $R_{ab}\,l^{a}n^{b}=-\Lambda$
imply that $A_{1}(x)$ and $A_{2}(y)$ are quadratic polynomials of $x$ and $y$
respectively, but in this case it turns out that the spacetime is flat.
Indeed, this can be grasped form the fact that the only non-vanishing Weyl
scalars in the general case are
\begin{equation}
\Psi_{2}^{+}\,=\,-\,\frac{\Lambda}{3}\quad\text{ and }\quad\Psi_{2}%
^{-}\,=\,-\,\frac{\Lambda}{3}\,,
\end{equation}
so that if $\Lambda=0$ then the Ricci tensor and the Weyl tensor are both
identically zero, which implies that the space is flat. Hence, let us just
consider the case of non-zero cosmological constant. The results of this
paragraph lead to the conclusion that the metric (\ref{LineEl.P1P2-cte}) with
$S$ being a non-zero constant and the functions $A_{1}$ and $A_{2}$ given by
(\ref{A1A2-P1P2cte}) is a solution of Einstein's vacuum equation with
cosmological constant $\Lambda$. Such solution turns out to admit the
following two Killing-Yano tensors:
\begin{equation}
\boldsymbol{Y}_{1}\,=\,\boldsymbol{l}\wedge\boldsymbol{n}\quad\;,\;\quad
\boldsymbol{Y}_{2}\,=\,-i\,\boldsymbol{m}_{1}\wedge\boldsymbol{m}_{2}\,.
\label{KY-P1P2S-cte}%
\end{equation}

A convenient choice of coordinates for the solution considered in the present
subsection is:
\begin{equation}
\hat{\tau}\,=\,\frac{\sqrt{s\,\Lambda\,a_{2}}}{p_{1}\,+\,p_{2}}\,(\tau
\,+\,p_{1}\sigma)\quad,\quad\hat{\sigma}\,=\,\frac{\sqrt{s\,\Lambda\,a_{1}}%
}{p_{1}\,+\,p_{2}}\,(\tau\,-\,p_{2}\sigma)
\end{equation}%
\begin{equation}
\hat{x}\,=\,x\,\sqrt{s\,\Lambda}\,+\,b_{1}\,-\,\frac{\pi}{2}\quad,\quad\hat
{y}\,=\,y\,\sqrt{s\,\Lambda}\,+\,b_{2}\,-\,\frac{\pi}{2}\,.
\end{equation}
With these coordinates, we conclude that the general solution of Einstein's
vacuum equation for the metric (\ref{LineEl.P1P2-cte}) with both functions
$S_{1}$ and $S_{2}$ being constant is given by:
\begin{equation}
ds^{2}\,=\,\frac{1}{\Lambda}\,\left[  \,-\,\sin^{2}(\hat{y})\,d\hat{\tau}%
^{2}\,+\,d\hat{y}^{2}\,+\,d\hat{x}^{2}\,+\,\sin^{2}(\hat{x})\,d\hat{\sigma
}^{2}\,\right]  \,. \label{Nariai}%
\end{equation}
The latter space is just the product of the 2-dimensional (Anti-)de Sitter
space with a sphere of radius $\Lambda^{-1/2}$, $(A)dS_{2}\times S^{2}$. This
space can be seen as a double Wick rotated version of the Nariai spacetime
\cite{Ortaggio-Nariai}. In terms of these new coordinates, the KY tensors of
Eq. (\ref{KY-P1P2S-cte}) are given by
\begin{equation}
\boldsymbol{Y}_{1}\,=\,\frac{\sin(\hat{y})}{\Lambda}\,d\hat{y}\wedge
d\hat{\tau}\quad\;,\;\quad\boldsymbol{Y}_{2}\,=\,\frac{\sin(\hat{x})}{\Lambda
}\,d\hat{x}\wedge d\hat{\sigma}\,. \label{KY-P1P2S-cte2}%
\end{equation}
Since the spaces $(A)dS_{2}$ and $S^{2}$ are maximally symmetric spaces of
dimension two it follows that they both admit three independent Killing
vectors. Therefore, the metric (\ref{Nariai}) should have six Killing vector
fields. Indeed, one can check that the following six 1-forms are independent
Killing fields:
\begin{align}
\boldsymbol{k}_{1}  &  \,=\,\sin^{2}(\hat{x})\,d\hat{\sigma}\,,\\
\boldsymbol{k}_{2}  &  \,=\,\sin(\hat{\sigma})\,d\hat{x}\,+\,\sin(\hat{x}%
)\cos(\hat{x})\cos(\hat{\sigma})\,d\hat{\sigma}\,,\\
\boldsymbol{k}_{3}  &  \,=\,\cos(\hat{\sigma})\,d\hat{x}\,-\,\sin(\hat{x}%
)\cos(\hat{x})\sin(\hat{\sigma})\,d\hat{\sigma}\,,\\
\boldsymbol{k}_{4}  &  \,=\,\sin^{2}(\hat{y})\,d\hat{\tau}\,,\\
\boldsymbol{k}_{5}  &  \,=\,\sinh(\hat{\tau})\,d\hat{y}\,+\,\sin(\hat{y}%
)\cos(\hat{y})\cosh(\hat{\tau})\,d\hat{\tau}\,,\\
\boldsymbol{k}_{6}  &  \,=\,\cosh(\hat{\tau})\,d\hat{y}\,+\,\sin(\hat{y}%
)\cos(\hat{y})\sinh(\hat{\tau})\,d\hat{\tau}\,.
\end{align}
It turns out that the Killing tensors generated by the square of the
Killing-Yano tensors $\boldsymbol{Y}_{1}$ and $\boldsymbol{Y}_{2}$ are
reducible, namely they can be written as linear combination of symmetrized
products of the Killing vectors. Indeed, defining $Q_{1\,ab}=Y_{1\,a}%
^{\phantom{1\,a}c}\,Y_{1\,cb}$ and $Q_{2\,ab}=Y_{2\,a}^{\phantom{2\,a}c}%
\,Y_{2\,cb}$, it is easy to check that
\begin{equation}
\boldsymbol{Q}_{1}=\frac{1}{2\Lambda}(\boldsymbol{k}_{6}\odot\boldsymbol{k}%
_{6}-\boldsymbol{k}_{5}\odot\boldsymbol{k}_{5}-\boldsymbol{k}_{4}%
\odot\boldsymbol{k}_{4})\quad\text{and}\quad\boldsymbol{Q}_{2}=-\,\frac
{1}{2\Lambda}(\boldsymbol{k}_{1}\odot\boldsymbol{k}_{1}+\boldsymbol{k}%
_{2}\odot\boldsymbol{k}_{2}+\boldsymbol{k}_{3}\odot\boldsymbol{k}_{3})\,.
\end{equation}

\section*{Acknowledgments}

C.B. thanks the Brazilian funding agency CAPES (Coordena\c{c}\~{a}o de
Aperfei\c{c}oamento de Pessoal de N\'{\i}vel Superior) for the financial
support. Research of A. A. is supported in part by FONDECYT grant 1141073 and
Newton-Picarte Grants DPI20140053 and DPI20140115. A.A. thanks Eloy Ayon-Beato
for his valuable introduction to the subject discussed in this paper. C.B.
thanks Professor Bruno Carneiro da Cunha for valuable discussions.



\begin{thebibliography}{99}                                                                                               %


\bibitem {Carter-KleinG}B. Carter, \textit{Hamilton-Jacobi and Schrodinger
separable solutions of Einstein's equations}, Commun. Math. Phys. \textbf{10}
(1968), 280.

\bibitem {Kubiznak:2008qp}D.~Kubiznak, \textit{Hidden Symmetries of
Higher-Dimensional Rotating Black Holes}, arXiv:0809.2452 [gr-qc].


\bibitem {BenentiFrancaviglia}S. Benenti and M. Francaviglia, \textit{Remarks
on certain separability structures and their applications to general
relativity}, Gen. Relativ. Gravit. \textbf{10} (1979), 79.

\bibitem {DeWitt:1973uma}C.~DeWitt and B.~S.~DeWitt, ``Proceedings, Ecole
d'Et\'{e} de Physique Th\'{e}orique: Les Astres Occlus : Les Houches, France,
August, 1972,''

\bibitem {Chandra-Dirac}S. Chandrasekhar, \textit{The solution of Dirac's
equation in Kerr geometry}, Proc. R. Soc. Lond. A. \textbf{349} (1976), 571;
B. Carter and R. McLenaghan, \textit{Generalized total angular momentum
operator for the Dirac equation in curved space-time}, Phys. Rev. D
\textbf{19} (1979), 1093.

\bibitem {Carter-constant}B. Carter, \textit{Global structure of the Kerr
family of gravitational fields}, Phys. Rev. \textbf{174} (1968), 1559.

\bibitem {Walk-Pen}M. Walker and R. Penrose, \textit{On quadratic first
integrals of the geodesic equations for type \{22\} spacetimes}, Commun. Math.
Phys. \textbf{18} (1970), 265.

\bibitem {Collinson-Stephani}C. Collinson, \textit{On the relationship between
Killing tensors and Killing-Yano tensors}, Int. J. Theor. Phys. \textbf{15}
(1976), 311; H. Stephani, \textit{A note on Killing tensors}, Gen. Relativ.
Gravit. \textbf{9} (1978), 789.

\bibitem {Frolov_KY}V. Frolov and D. Kubiz\v{n}\'{a}k,
\textit{Higher-dimensional black holes: hidden symmetries and separation of
variables}, Class. Quant. Grav. \textbf{25} (2008), 154005.

\bibitem {Yasui}Y. Yasui and T. Houri, \textit{Hidden symmetry and exact
solutions in Einstein gravity}, Prog. Theor. Phys. Suppl. \textbf{189} (2011), 126.

\bibitem {Kubiz}D. Page \textit{et al.}, \textit{Complete integrability of
geodesic motion in general Kerr-NUT-AdS spacetimes}, Phys. Rev. Lett.
\textbf{98} (2007), 061102.

\bibitem {Krtous}P. Krtou\v{s} \textit{et al.}, \textit{Killing-Yano tensors,
rank-2 Killing tensors, and conserved quantities in higher dimensions}, JHEP
\textbf{0702} (2007), 004.

\bibitem {Frol-KG}V. Frolov, P. Krtou\v{s} and D. Kubiz\v{n}\'{a}k,
\textit{Separability of Hamilton-Jacobi and Klein-Gordon equations in general
Kerr-NUT-AdS spacetimes}, JHEP \textbf{0702} (2007), 005.

\bibitem {Oota}T. Oota and Y. Yasui, \textit{Separability of Dirac equation in
higher dimensional Kerr-NUT-de Sitter spacetime}, Phys. Lett. B \textbf{659}
(2008), 688.

\bibitem {Teukolsky}S. Teukolsky, \textit{Rotating black holes: separable wave
equations for gravitational and electromagnetic perturbations}, Phys. Rev.
Lett. \textbf{29} (1972), 1114.

\bibitem {OotaPerturb}T. Oota and Y. Yasui, \textit{Separability of
gravitational perturbation in generalized Kerr-NUT-de Sitter spacetime}, Int.
J. Mod. Phys. A \textbf{25} (2010) 3055.

\bibitem {KY-SUSY}G. Gibbons, R. Rietdijk and J. van Holten, \textit{SUSY in
the sky}, Nucl. Phys. B \textbf{404} (1993), 42; M. Tanimoto, \textit{The role
of Killing-Yano tensors in supersymmetric mechanics on a curved manifold},
Nucl. Phys. B \textbf{442} (1995), 549

\bibitem {Chong:2004na}Z.-W.~Chong, M.~Cvetic, H.~Lu and C.~N.~Pope,
\textit{Charged rotating black holes in four-dimensional gauged and ungauged
supergravities}, Nucl.\ Phys.\ B \textbf{717} (2005) 246
doi:10.1016/j.nuclphysb.2005.03.034 [hep-th/0411045].

\bibitem {AlonsoAlberca:2000cs}N.~Alonso-Alberca, P.~Meessen and T.~Ortin,
``Supersymmetry of topological Kerr-Newman-Taub-NUT-AdS space-times,''
Class.\ Quant.\ Grav.\ \textbf{17} (2000) 2783 doi:10.1088/0264-9381/17/14/312 [hep-th/0003071].

\bibitem {Klemm:2013eca}D.~Klemm and M.~Nozawa, ``Supersymmetry of the
C-metric and the general Plebanski-Demianski solution,'' JHEP \textbf{1305}
(2013) 123 doi:10.1007/JHEP05(2013)123 [arXiv:1303.3119 [hep-th]].

\bibitem {Chow:2010sf}D.~D.~K.~Chow, ``Single-charge rotating black holes in
four-dimensional gauged supergravity,'' Class.\ Quant.\ Grav.\ \textbf{28}
(2011) 032001 doi:10.1088/0264-9381/28/3/032001 [arXiv:1011.2202 [hep-th]].

\bibitem {Chow:2010fw}D.~D.~K.~Chow, ``Two-charge rotating black holes in
four-dimensional gauged supergravity,'' Class.\ Quant.\ Grav.\ \textbf{28}
(2011) 175004 doi:10.1088/0264-9381/28/17/175004 [arXiv:1012.1851 [hep-th]].

\bibitem {Chow:2013tia}D.~D.~K.~Chow and G.~Comp\`{e}re, ``Seed for general
rotating non-extremal black holes of $\mathcal{N}= 8$ supergravity,''
Class.\ Quant.\ Grav.\ \textbf{31} (2014) 022001
doi:10.1088/0264-9381/31/2/022001 [arXiv:1310.1925 [hep-th]].

\bibitem {Chow:2013gba}D.~D.~K.~Chow and G.~Comp\`{e}re, ``Dyonic AdS black
holes in maximal gauged supergravity,'' Phys.\ Rev.\ D \textbf{89} (2014) 6,
065003 doi:10.1103/PhysRevD.89.065003 [arXiv:1311.1204 [hep-th]].

\bibitem {Chow:2014cca}D.~D.~K.~Chow and G.~Comp\`{e}re, ``Black holes in N=8
supergravity from SO(4,4) hidden symmetries,'' Phys.\ Rev.\ D \textbf{90}
(2014) 2, 025029 doi:10.1103/PhysRevD.90.025029 [arXiv:1404.2602 [hep-th]].

\bibitem {Anabalon:2012ta}A.~Anabalon, ``Exact Black Holes and Universality in
the Backreaction of non-linear Sigma Models with a potential in (A)dS4,'' JHEP
\textbf{1206} (2012) 127 doi:10.1007/JHEP06(2012)127 [arXiv:1204.2720 [hep-th]].

\bibitem {Robinson:2004zz}D.~Robinson, ``Four decades of black holes
uniqueness theorems,''


\bibitem {RobinsonManifolds}P. Nurowski and A. Trautman, \textit{Robinson
manifolds as the Lorentzian analogs of Hermite Manifolds}, Differential
Geometry and its Applications \textbf{17}(2002), 175.

\bibitem {Goldberg-Sachs}J. Goldberg and R. Sachs, \textit{A theorem on Petrov
types}, Gen. Relativ. Gravit. \textbf{41} (2009), 433. Republication of the
original 1962 paper.

\bibitem {GS-Conformal}I. Robinson and A. Schild, \textit{Generalization of a
theorem by Goldberg and Sachs}, J. Math. Phys. \textbf{4} (1963), 484.

\bibitem {Plebanski2}J. F. Pleba\'{n}ski and S. Hacyan, \textit{Null geodesic
surfaces and Goldberg-Sachs theorem in complex Riemannian spaces}, J. Math.
Phys. \textbf{16} (1975), 2403.

\bibitem {Bat1}C. Batista, \textit{Weyl tensor classification in
four-dimensional manifolds of all signatures}, Gen. Relativ. Gravit.
\textbf{45} (2013), 785.

\bibitem {FrolovBook}V. Frolov and A. Zelnikov, \textit{Introduction to black
hole physics}, Oxford University Press (2011).

\bibitem {Santillan}O. Santillan, \textit{Hidden symmetries and supergravity
solutions}, J. Math. Phys. \textbf{53} (2012), 043509; M. Cariglia,
\textit{Quantum mechanics of Yano tensors: Dirac equation in curved
spacetime}, Class. Quant. Grav. \textbf{21} (2004), 1051.

\bibitem {Dietz}W. Dietz and R. Rudiger , \textit{Space-times admitting
Killing-Yano tensors I}, Proc. R. Soc. Lond. A \textbf{375} (1981), 361; W.
Dietz and R. Rudiger, \textit{Space-times admitting Killing-Yano tensors II},
Proc. R. Soc. Lond. A \textbf{381} (1982), 315.

\bibitem {Hall-KY}G. Hall, \textit{Killing-Yano tensors in general
relativity}, Int. J. Theor. Phys. \textbf{26} (1987), 71.

\bibitem {Bat-KYn-1}C. Batista, \textit{Killing-Yano tensors of order n-1},
Class. Quant. Grav. \textbf{31} (2014), 165019.

\bibitem {Bat-Book-art2}C. Batista, \textit{Generalizing the Petrov
Classification}, Lambert Academic Publishing (2014); C. Batista, \textit{A
generalization of the Goldberg-Sachs theorem and its consequences}, Gen.
Relativ. Gravit. \textbf{45} (2013), 1411.

\bibitem {ParaComplex}S. Ivanov and S. Zamkovoy, \textit{ParaHermitian and
paraquaternionic manifolds}, Differ. Geom. Appl. \textbf{23} (2005), 205; V.
Cruceanu, P. Fortuny, and P.M. Gadea, \textit{A survey on paracomplex
geometry}, Rocky Mountain J. Math. \textbf{26} (1996), 83.

\bibitem {Ortaggio-Nariai}M. Ortaggio, \textit{Impulsive waves in the Nariai
universe}, Phys. Rev. D \textbf{65} (2002), 084046.

\bibitem {GrifPodol-Book}J.~B.~Griffiths and J.~Podolsky, \textit{ Exact
Space-Times in Einstein's General Relativity}, Cabridge University Press (2009).
\end{thebibliography}
\end{document}